\documentclass[twocolumn,showkeys,showpacs,preprintnumbers,prd,superscriptaddress,nofootinbib]{revtex4-1}
\usepackage{graphicx}
\usepackage{epsf}
\usepackage{bm}
\usepackage{amsmath}
\usepackage{amsfonts}
\usepackage{amssymb}
\usepackage{epstopdf}
\usepackage{hyperref}
\usepackage{color}
\usepackage{verbatim}
\usepackage{multirow}
\usepackage{bm}
\usepackage{hyperref}

\begin{document}

\title{Revisiting the temperature evolution law of the CMB with gaussian processes}

\author{Felipe Avila}
\email{felipeavila@on.br}
\affiliation{Observat\'orio Nacional, Rua General Jos\'e Cristino 77, 
S\~ao Crist\'ov\~ao, 20921-400 Rio de Janeiro, RJ, Brazil}

\author{Alexander Bonilla Rivera}
\email{alex.acidjazz@gmail.com}
\affiliation{Instituto de F\'{i}sica, Universidade Federal Fluminense, 24210-346 Niter\'{o}i, RJ, Brazil}

\author{Rafael C. Nunes}
\email{rafadcnunes@gmail.com}
\affiliation{Instituto de F\'{i}sica, Universidade Federal do Rio Grande do Sul, 91501-970 Porto Alegre RS, Brazil}
\affiliation{Divis\~{a}o de Astrof\'{i}sica, Instituto Nacional de Pesquisas Espaciais, Avenida dos Astronautas 1758, S\~{a}o Jos\'{e} dos Campos, 12227-010, S\~{a}o Paulo, Brazil}

\author{R. F. L. Holanda}
\email{holandarfl@gmail.com}
\affiliation{Universidade Federal do Rio Grande do Norte, Departamento de F\'{i}sica Te\'{o}rica e Experimental, 59300-000, Natal - RN, Brazil}
\affiliation{Departamento de F\'{\i}sica, Universidade Federal de Campina Grande, 58429-900, Campina Grande - PB, Brasil}

\author{Armando Bernui}
\email{bernui@on.br}
\affiliation{Observat\'orio Nacional, Rua General Jos\'e Cristino 77, 
S\~ao Crist\'ov\~ao, 20921-400 Rio de Janeiro, RJ, Brazil}

\begin{abstract}
In this work, we perform a statistical inference of the classical background law governing the evolution of the temperature of the cosmic microwave background radiation (CMB), given by $T_{\rm CMB}(z) = T_0(1 + z)$. To this end, we employ Gaussian Process (GP) regression techniques to reconstruct the temperature evolution based on two observational datasets: (i) CMB–Sunyaev-Zel’dovich (SZ) cluster measurements and (ii) CMB–interstellar medium (ISM) interaction data. Our analysis reveals interesting results that may suggest potential deviations from the standard temperature-redshift relation, particularly at low redshifts ($z < 0.5$), where discrepancies up to $\sim$2$\sigma$ are observed. Additionally, we identify a mild but noteworthy tension, also at the $\sim$2$\sigma$ level, between our GP inferred value of the present-day CMB temperature, $T_{\rm CMB}(z=0)$, and the precise direct measurement from the COBE/FIRAS experiment. We also explore possible phenomenological implications of our findings, including interpretations associated with possible variations in fundamental constants, such as the fine-structure constant $\alpha$, which could provide a physical explanation for the observed deviations at low redshift.
\end{abstract}

\keywords{CMB, Cosmology, Gaussian Process}

\pacs{}

\maketitle

\section{Introduction}
\label{sec:introduction}

The cosmic microwave background (CMB) radiation is one of the most essential and reliable tools for studying the dynamics, composition, and evolution of the Universe~\citep{Hu02,Dodelson_book}. According to the predictions of the hot Big Bang model, the energy of CMB photons decreases as the Universe expands, following the temperature--redshift relation (TRR) 
\begin{equation}
T_{\text{CMB}}(z) = T_0 (1+z) \,,
\end{equation}
where \( T_0 \) is the present-day thermodynamic temperature of the CMB. Accurately determining \( T_0 \) is crucial, as it serves as a fixed input parameter in the analysis of high-precision CMB data from space missions such as \textit{Planck}~\citep{Planck20}, ACT~\citep{ACT:2025fju}, and SPT~\citep{Saro14}. A precise value of \( T_0 \) is also required for the proper calibration of temperature maps, including the accurate modeling and subtraction of the dipole component from the raw data, considering that masks need to be applied due to foregrounds and that CMB temperature fluctuations may exhibit deviations from 
Gaussianity~\citep{Bernui2010,Bernui2009,BMRT2007IJMPD,Gruppuso2014,Planck-iso+stat2020}. 

Since the mid-20\textsuperscript{th} century, various methods and instruments have been employed to measure \( T_0 \). These include observations from COBE/FIRAS~\citep{Smoot1987,Mather94,Fixen96}, molecular line spectroscopy~\citep{Klimenko20}, and balloon-borne experiments (see reviews by~\citep{Bersanelli2002,Fixen09}). The currently accepted value, measured with remarkable precision, is
\begin{equation}
T_0 = 2.72548 \pm 0.00057~\text{K} \,,
\end{equation}
as reported by~\cite{Fixen09}. 

Previous studies have explored several aspects related to the importance of an accurate measurement of \( T_0 \)~\citep{Fixen09,Fixen96}. Some have investigated how uncertainties in \( T_0 \) can impact the cosmological parameters derived from CMB temperature fluctuation data~\citep{Ivanov20}. Other works have analyzed how combining current CMB observations with external datasets can lead to more precise determinations of \( T_0 \)~\citep{Gush90,Noterdaeme11}. 
Moreover, correlations between \( T_0 \) and measurements of the Hubble constant have also been examined~\citep{Ivanov20}. In fact, analogously to the redshift drift phenomenon~\citep{Guo16}, measuring the cooling of the CMB provides a potential avenue to constrain the Hubble constant~\citep{Abitbol20}. Further research has also assessed whether future CMB anisotropy experiments could independently determine \( T_0 \) with high precision, as well as the possibility of such measurements in other astrophysical contexts~\citep{Ivanov20}.

On the other hand, several extensions of the standard cosmological model, $\Lambda$CDM, predict deviations from the standard TRR. These are often tested using phenomenological parameterizations 
\begin{equation}
\label{main_eq}
T_{\rm CMB}(z) = T_0 (1 + z)^{1 - \beta} \,,
\end{equation}
as proposed in~\citep{Luzzi2009,Noterdaeme11,Saro14,Luzzi15,Martino15,Klimenko20,Arjona20,Li20,Riechers22}, or through more specific relations derived from particular theoretical frameworks. Such frameworks may involve variations in fundamental physical constants, including the gravitational constant and the fine-structure constant, among other possibilities~\citep{Noterdaeme11,Saro14,Luzzi15,Martino15,Klimenko20,Arjona20,Li20,Riechers22,Lima00}.

These investigations underscore the cosmological relevance of a precise determination of \( T_0 \), which influences not only our interpretation of the CMB but also our understanding of the Universe evolution. As observational precision improves, \( T_0 \) remains a key parameter in testing both the standard model and its extensions. For instance, Ref.~\cite{2014PhRvD..90l4064H} presents a framework based on scalar-tensor theories that violate the Einstein Equivalence Principle in the electromagnetic sector. In this scenario, matter fields couple to a scalar field, modifying the electromagnetic sector and leading to observable effects such as violations of the Cosmic Distance Duality Relation (CDDR), variations in the fine-structure constant \( \Delta\alpha/\alpha \), and deviations in the redshift evolution of the CMB temperature (see also~\cite{2024EPJC...84.1120F} for recent constraints).

In this work, we propose updated constraints on \( T_{\rm CMB,0} \) and \( \beta \) using the CMB temperature evolution data set \( T_{\rm CMB}(z) \), for $z \in [0.037, 3.2874]$. 
Currently, there are essentially two methods to measure \( T_{\rm CMB}(z) \). 
The first one relies on distortions in the observed blackbody spectrum of the CMB caused by the Sunyaev-Zel'dovich (SZ) effect. 
This method is practically limited to low redshifts (\( z < 0.6 \)) 
due to the scarcity of measurements of this effect at higher 
redshifts~\citep{Fabri78,Rephaeli80,Battistelli_2002,Luzzi2009}.
The second method exploits the fact that, in the interstellar medium, atomic or molecular species can reach radiative equilibrium with CMB photons. 
In such environments, the excitation temperature of specific transitions provides a direct probe of the CMB temperature at high redshifts~\citep{Bahcall1968,Riechers22}.

Our methodology is based on the statistical reconstruction of the function \( T_{\rm CMB}(z) \) using Gaussian Processes (GP), a non-parametric supervised machine learning technique~\citep{GP}. With this approach, we reconstruct not only the function \( T_{\rm CMB}(z) \), but also its first and second derivatives, \( T'_{\rm CMB}(z) \) and \( T''_{\rm CMB}(z) \), and the redshift-dependent parameter \( \beta(z) \), across the full redshift range covered by the data, i.e., \( z \in [0.037, 3.2874] \). 
To further investigate the implications of a non-zero \( \beta \), we also reconstruct the function describing the variation of the fine-structure constant, in the form \( \Delta\alpha/\alpha \).

It is important to emphasize that our work differs significantly from previous studies~\citep{Arjona20,Bengaly20} in several key aspects:  
\begin{itemize}
    \item[(i)] Data: the three temperature measurement compilations share many common points, but they are not identical. Our analysis shares a similar goal with that of~\citep{Arjona20} (hereafter A20), but we obtain different results for \( T_0 \) and \( \beta_0 \equiv \beta(z=0) \);  
    \item[(ii)] Methodology:~\citep{Bengaly20} (hereafter B20) employs both parametric (\( \chi^2 \) fitting) and non-parametric (GP) techniques but focuses solely on constraining \( T_0 \). A20, in contrast, adopts a purely non-parametric approach based on a genetic algorithm, which is methodologically distinct from GP.
\end{itemize}
When presenting our results, we compare them with those of A20 and B20 to identify potential inconsistencies and biases stemming from differences in the adopted temperature datasets and reconstruction methods. Finally, as previously mentioned, we will also derive new constraints on possible variations of the fine-structure constant within the class of theories discussed in~\cite{2014PhRvD..90l4064H}.

The structure of this paper is as follows. In Section~\ref{sec:data_method}, we provide a brief overview of the temperature measurements compiled by~\citep{Riechers22}, highlighting specific aspects of the data. We also introduce the fundamental equations used in our analysis, including error propagation, and describe the non-parametric reconstruction technique employed, namely GP. Section~\ref{sec:Results} presents our main results, including the reconstructions of \( T_{\rm CMB}(z) \), its first and second derivatives, \( T'_{\rm CMB}(z) \) and \( T''_{\rm CMB}(z) \), as well as \( \beta(z) \), along with a discussion of their cosmological implications.
In Section~\ref{sec:Results2}, we explore phenomenological models based on the reconstructed functions \( \beta(z) \) and \( \Delta\alpha/\alpha \). Finally, Section~\ref{final} provides our concluding remarks.

\section{Data-set and Methodology}
\label{sec:data_method}

In this section, we present the details of our dataset and the statistical methodology employed in our analysis. We extracted 49 data points from~\cite{Riechers22}, covering the redshift interval \( z \in [0.037, 3.2874] \). It is worth noting that these measurements were obtained using two methodologies, as described below:

\begin{itemize}
    \item \textbf{CMB-Sunyaev-Zel'dovich (SZ) samples:} The data at low redshifts uses the well-known thermal Sunyaev-Zel'dovich effect (tSZ), which is a distortion of the CMB black-body spectrum caused by inverse Compton scattering. In~\cite{Riechers22}, three low-redshift samples are provided. The first~\citep{Hurier14}, using Planck clusters~\citep{Planck14}, measured 18 data points of \( T(z) \) in the redshift range \( 0.037 \leq z \leq 0.972 \). The second~\citep{Martino15}, also using Planck data, measured 6 data points of \( T_{\rm CMB}(z) \) in the redshift range \( 0.042 \leq z \leq 0.274 \). The last sample~\citep{Saro14} measured 12 data points of \( T(z) \) in the redshift range \( 0.129 \leq z \leq 1.022 \) using clusters from the South Pole Telescope~\citep{Reichardt13}. The key difference between these 36 \( T_{\rm CMB}(z) \) data points lies in the number of clusters included within the redshift bins. In~\cite{Riechers22}, the methodology for each data point is specified.

    \item \textbf{CMB-Interstellar Medium Interaction (IMI):} For high redshifts, the data are obtained by measuring the atomic and molecular excitation temperatures in a diffuse gas present in absorption lines of quasar spectra. These populations are in radiative equilibrium with the CMB radiation, and the additional background temperature provides measurements of the evolution of \( T_{\rm CMB}(z) \) at redshifts between approximately 1.7 and 3.3, yielding 13 additional data points (refer to Table 1 in~\cite{Riechers22}).
\end{itemize}

Our final sample contains 49 data points from~\cite{Riechers22}, excluding 5 data points that provide only upper limits, the starburst galaxy HFLS3, and the data point at \( z = 0 \) (with \( T_0 = 2.72548 \pm 0.00057 \)), for the convenience of our statistical analysis using GP. Note that the sample we selected includes 12 additional data points compared to A20, which improves the statistical sample by approximately 24\%.

\subsection{Minimal Theoretical Inputs}
\label{sec:equations}

A parameterization commonly adopted in the literature to test deviations in the temperature-redshift relation (TRR) is~\citep{Lima00}, given by equation (\ref{main_eq}), where \(\beta\) can either be a constant or exhibit a dependence on cosmic time. 
We can infer \(\beta(z)\) by taking the first derivative of equation (\ref{main_eq}):
\begin{equation}
\beta = 1 - (1+z)\frac{T'(z)}{T(z)} \,. 
\end{equation}
The uncertainty on \( \beta \) is calculated using standard error propagation techniques, which account for the statistical uncertainties inherent in the data. 
Under the assumption that there is no deviation from the standard TRR, we expect the first derivative of the CMB temperature, \( T'(z) \), to be constant and equal to \( T_0 \), i.e., \( T'(z) = T_0 \). 
This implies that the rate of change of the CMB temperature with respect to redshift should follow a simple, linear evolution. 
Additionally, the second derivative, \( T''(z) \), is expected to be zero, indicating no curvature in the TRR. 
In order to identify potential deviations from the TRR, it is crucial to study both the first and second derivatives of the reconstructed CMB temperature, \( T(z) \). 
These derivatives contain valuable cosmological information, as any significant deviation from the expected behavior would manifest in either the rate of change (first derivative) or the curvature (second derivative) of the temperature time evolution. 
By analyzing these derivatives, we can gain insights into possible modifications to the standard cosmological model, particularly in the context of new physics or alternative gravity theories that may influence the CMB temperature evolution.


\subsection{Gaussian Processes}
\label{sec:GP}

From a statistical perspective, Gaussian Processes (GPs) have been widely used in cosmology~\cite{Seikel_2012,Shafieloo12,seikel13,Zhang16,Busti14,Sahni14,Belgacem20,Pinho18,Cai17,Haridasu18,Zhang18,Wang17,Bengaly20,Arjona20,Sharma22}. 
GP allow us to extract crucial information about the dynamics of the Universe without assuming a specific parameterization or physical model for the dark sector~\cite{Bonilla20,Saboga24,GP,Oliveira23,Avila22,Avila22a,Bonilla21}. 
Based on observed data, GP reconstruct the function \( f(z) \) that best fits the data and propagate the statistical errors 
\begin{equation}
\label{eqn:GPs}
f(z_i) = \mathcal{GP}(\mu(z_i), \textrm{cov}[f(z_i), f(z_i)]) \,,
\end{equation}
ensuring the necessary statistical confidence for the analysis. 
GPs assume a point-to-point Gaussian distribution in the dataset of interest, providing a robust statistical framework for cosmological studies. 
The functions at different points \( z_i \) and \( z_{i+1} \) are correlated by a covariance matrix, or kernel, \( \textrm{cov}[f(z_i),f(z_i)] = k(z_i, z_{i+1}) \), which depends solely on a set of hyperparameters \( l \) and \( \sigma_f \) in the statistical sense. These hyperparameters describe the strength and extent of correlations between the reconstructed data points. Thus, the kernel for the Squared Exponential (SE) function is defined as
\begin{equation}
\label{eqn:kSE}
k_{SE}(z_i, z_{i+1}) = \sigma_f^2 \exp\left(-\frac{|z_i - z_{i+1}|^2}{2\, l^2}\right) \,.
\end{equation}
Here, \( l \) represents the coherence length of the correlation along the \( z \)-axis, and \( \sigma_f \) denotes the amplitude of the correlation in the ordinate direction. These hyperparameters typically remain constant, ensuring a good fit to the observed data, rather than mimicking specific behaviors of a model. The Gaussian Process (GP) optimizes these parameters with respect to the observed data and provides Bayesian-determined statistical weights.

The GP method is model-independent, meaning it does not assume a particular physical model, such as a cosmological model. Instead, it relies on a statistical kernel that defines correlations between reconstructed data points. Examples of such kernels include the Squared Exponential (SE) kernel or the Matérn kernels with parameters \( \nu = 5/2, 7/2, \) or \( 9/2 \), whose general functional form is given by

\begin{equation}
    k_{M_{\nu}}(\tau) = \sigma_f^2 \frac{2^{1 - \nu}}{\Gamma(\nu)} \left( \frac{\sqrt{2 \nu} \tau}{l} \right)^{\nu} K_{\nu}\left( \frac{\sqrt{2 \nu} \tau}{l} \right),
\end{equation}
where \( K_{\nu} \) is the modified Bessel function of the second kind, and \( \Gamma(\nu) \) is the Gamma function. The parameter \( \nu \) is strictly positive, and the SE kernel corresponds to the limit \( \nu \to \infty \) of the Matérn kernel.

For the statistical reconstructions in our work using the GP technique, we utilize the publicly available code from \cite{Seikel_2012} \footnote{\url{https://github.com/JCGoran/GaPP/tree/feature/python3}}. In the following sections, we present our main results.

\begin{table*}
\hspace{-1.0cm}
\begin{tabular}{|l|l|l|l|l|l|l|l|}
\hline
 &
  Squared Exponential  &
  Matérn ($\nu = 5/2$) &
  Matérn ($\nu = 7/2$) &
  Matérn ($\nu = 9/2$) &
  Rational Quadratic \\ \hline
$T_{0}$  &
  $2.73987 \pm 0.00793$ &
  $2.75562 \pm 0.01166$ &
  $2.74012 \pm 0.00802$ &
  $2.74004 \pm 0.00797$ &
  $2.73982 \pm 0.00792$ \\ \hline
$T_{0}'$ &
  $2.64943 \pm 0.04960$ &
  $2.38009 \pm 0.16578$ &
  $2.64656 \pm 0.05265$ &
  $2.64762 \pm 0.05115$ &
  $2.64854 \pm 0.04975$ \\ \hline
$T_{0}''$ &
  $0.23006 \pm 0.12793$ &
  $1.82423 \pm 1.42434$ &
  $0.24305 \pm 0.16698$ &
  $0.23755 \pm 0.14838$ &
  $0.22560 \pm 0.12691$ \\ \hline
$\beta_{0}$ &
  $0.03301 \pm 0.01954$ &
  $0.13628 \pm 0.07923$ &
  $0.03415 \pm 0.02077$ &
  $0.03373 \pm 0.02017$ &
  $0.03331 \pm 0.01961$ \\ \hline
\end{tabular}
\caption{Inference at the 1$\sigma$ level for the quantities \( T_{\rm CMB}(z=0) \), \( T'_{\rm CMB}(z=0) \), \( T''_{\rm CMB}(z=0) \), and \( \beta(z=0) \) is performed for different kernel functions used in the GP reconstruction.}
\label{tab:results}
\end{table*}


\section{Main Results}
\label{sec:Results}


Our main results are the non-parametric reconstruction of $T(z)$, $T'(z)$, $T''(z)$, and $\beta(z)$, using the sample of 49 
temperature data points $\{ T_{\rm CMB}(z_i) \}$ presented in Section~\ref{sec:data_method}. 
Additionally, we tested different reconstruction kernels to assess the robustness of our approach. Ideally, a reliable reconstruction should be independent of both the chosen kernel and the number of data points\footnote{A high-quality reconstruction does not require thousands of data points. Rather, it depends on a dataset with minimal fluctuations across the redshift range.}, allowing us to evaluate the quality of the dataset based on the consistency of the results.

\begin{figure*}[htbp]
\centering
\includegraphics[scale=0.6]{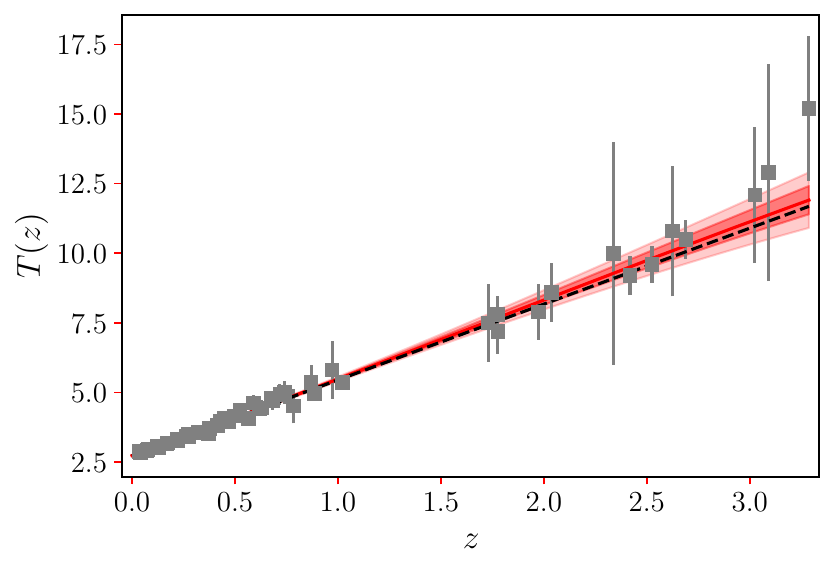} \,\,\,\,
\includegraphics[scale=0.6]{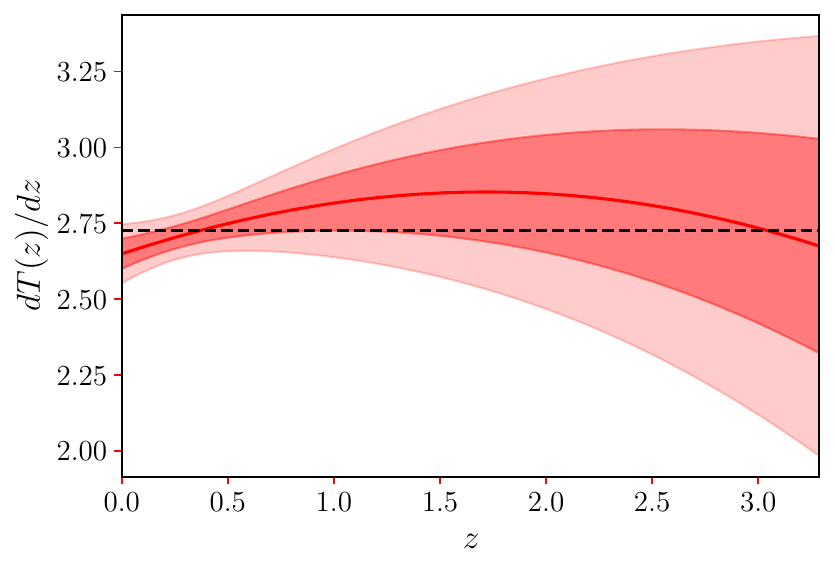}
\caption{Left panel: GP reconstruction of \( T(z) \) using the SE kernel. Right panel: Same as in the left panel, but for the quantities \( T_{\rm CMB}'(z) \). The black dashed line correspond to the expect value.}
\label{fig:Tz_recon}
\end{figure*}

\begin{figure*}[htbp]
\centering
\includegraphics[scale=0.6]{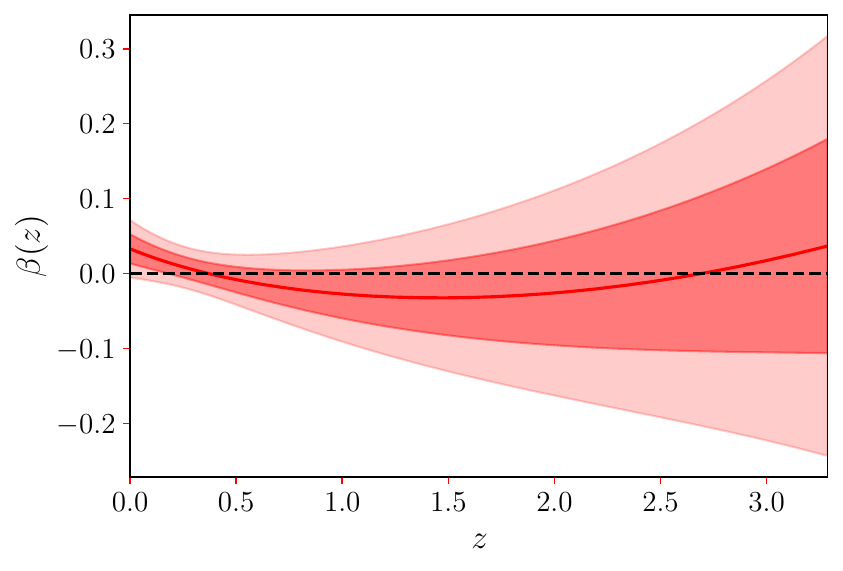} \,\,\,\,
\includegraphics[scale=0.6]{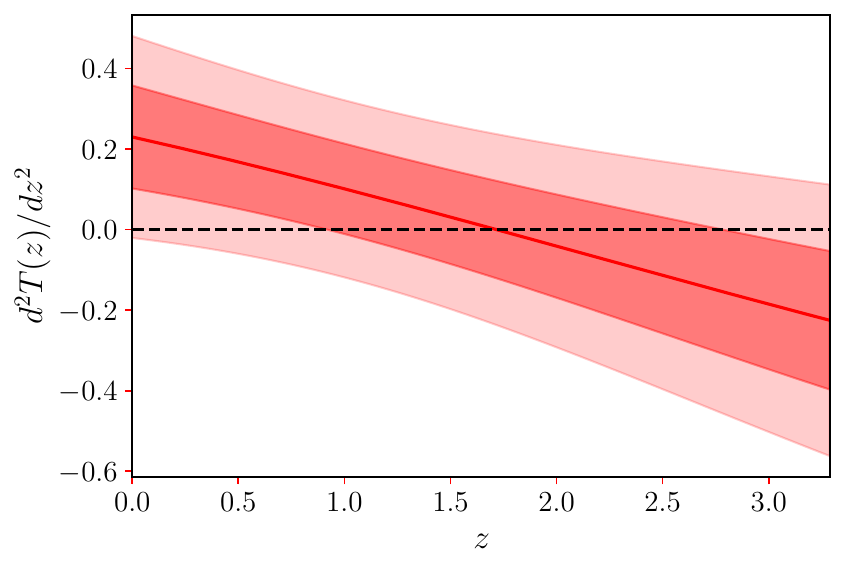}
\caption{Left panel: GP reconstruction of $\beta(z)$ using the SE kernel. Right panel: Same as the left panel, but for the function $T_{\rm CMB}''(z)$. The black dashed line correspond to the expect value.}
\label{fig:beta_recon}
\end{figure*}

Our analyses for GP reconstruction use several kernels, finding in 
each case the corresponding value \( T_{\rm CMB}(z=0) \), the results 
are shown in Table \ref{tab:results}. 
These reconstructions utilize different kernels to assess the robustness and reliability of our results. Upon analyzing the full sample, we observe that, with the exception of the Matérn kernel with \( \nu = 5/2 \), all other kernel-based reconstructions are consistent with the TRR at the \( 2\sigma \) confidence level (CL). It is crucial to highlight that the Matérn \( \nu = 5/2 \) kernel is highly sensitive to fluctuations in the data, which can lead to unstable results. As we increase the parameter \( \nu \), the Matérn kernel's behavior gradually converges toward that of the Squared Exponential (SE) kernel, which tends to produce smoother reconstructions. Thus, the result obtained with the \( \nu = 5/2 \) Matérn kernel is less reliable, given the inherent noise and sparsity of the dataset. 

It is interesting to note that our inference for the value of $T_0$ is slightly higher than the direct measurements obtained by COBE/FIRAS~\citep{Smoot1987,Mather94,Fixen96}. 
In~\citep{Rasanen:2015kca}, the authors assume the same deviation from the standard temperature evolution law as adopted in this work, but apply it to CMB anisotropy data from Planck (TT, TE, EE + lowP only). 
In their analysis, a positive correlation between $\beta$ and the Hubble constant $H_0$ is observed. 
Interpreted physically, this suggests that $\beta$ is positively correlated with the overall expansion rate of the universe. 
In this context, one can argue that $\beta$, $T_0$, and $H_0$ may all be positively correlated. 
A possible explanation for this trend is provided in the physical scenario originally proposed in~\citep{Kumar:2018yhh}, in which dark matter (DM) is coupled to photons, leading to non-conservation of particle number for both species. 
This interaction can naturally induce a positive correlation with $H_0$, where the dimensionless parameter characterizing the strength of the DM-photon coupling, $\Gamma_\gamma$, can be mapped as $\Gamma_\gamma \propto \beta$. 
Therefore, among other possible interpretations, this physical mechanism may accommodate slightly elevated values of $T_0$.


To visually illustrate our findings, Figure \ref{fig:Tz_recon} presents the statistical reconstruction of the function $T_{\rm CMB}(z)$, shown with 1$\sigma$ and 2$\sigma$ confidence intervals in the left panel, and its associated temporal variation in the right panel. For better comparison, both panels include a dashed line representing the standard $\Lambda$CDM prediction, with the present-day CMB temperature $T_0$, fixed according to the COBE/FIRAS estimate.
Our statistical predictions exhibit good agreement with the expected $\Lambda$CDM evolution over cosmic time, within 2$\sigma$ confidence level. 
However, it is important to emphasize that our inference at the present epoch, $z = 0$ (as already highlighted in Table \ref{tab:results}), may introduce some tension when compared with direct measurements of $T_{\rm CMB}(z=0)$. This potential discrepancy deserves further investigation, particularly in light of its implications for extensions of the standard cosmological model.

Figure \ref{fig:beta_recon} also shows the reconstruction of the function $\beta(z)$, including the 1$\sigma$ and 2$\sigma$ confidence intervals. 
This function quantifies potential phenomenological deviations from the standard temperature evolution law. 
For redshifts $z < 0.5$, we observe a possible indication of $\beta > 0$, reaching nearly the 2$\sigma$ level at $z = 0$. 
A statistically similar behavior is observed for the second derivative of the CMB temperature, $T_{\rm CMB}''(z)$. 
These results may suggest that at low redshifts, that is, during the late-time cosmic evolution, there could be deviations from the standard $\Lambda$CDM cosmology, as inferred from the data analyzed here. This interpretation aligns with our previous discussion on the inferred values of $T_{\rm CMB}(z = 0)$: since the deviations appear to grow as $z$ approaches zero, they could potentially explain the tension between our reconstructed $T_{\rm CMB}(z=0)$ and the values obtained through direct measurements, as discussed above.

When comparing our results with those presented in A20 and B20, 
we can notice small differences in data samples and methodological choices that can lead to different results. 
In particular, our estimate of the present-day CMB temperature, $T_0$, agrees with the value reported by B20. In contrast, A20 does not perform an independent inference of $T_0$, instead fixing it to the COBE/FIRAS value of $T_0 = 2.72548$ K. 
With respect to the reconstruction of the deviation parameter $\beta(z)$, B20 does not provide any such analysis, which limits the possibility of a direct comparison. A20, on the other hand, reports a normalized version of $\beta(z)$, making a direct quantitative comparison with our results more challenging. Nonetheless, qualitatively, the $\beta(z)$ curve presented in A20 (see their Figure 1, right panel) appears consistent with the TRR within the 1$\sigma$ confidence level. Interestingly, although the amplitude of deviations is at the percent level in both the A20 and our compilations, the concavity of the reconstructed $\beta(z)$ function in A20 differs from ours. This suggests a possible dependence on the methodological approach used in each analysis—namely, Gaussian Processes in our case and Genetic Algorithms in A20.

When a parameter is constrained at a specific redshift, typically at $z = 0$, GP allows its inference to be extended across the entire redshift range. This flexibility is particularly advantageous when dealing with smooth and well-behaved functions, such as the one given in equation~(1).  As a result, we can examine potential deviations of our reconstructed function $T_{\rm CMB}(z)$ from the standard TRR. 

To quantify these deviations, we define a tension function, $\mathcal{T}(z)$, as follows~\citep{Rocco23} 
\begin{equation} \mathcal{T}(z) = \frac{|T_{\rm GP}(z) - T_{\rm COBE}(z)|}{\sqrt{\sigma_{\rm GP}^{2}(z) + \sigma_{\rm COBE}^{2}(z)}} \,, 
\end{equation} 
where $T_{\rm GP}(z)$ denotes the temperature reconstructed via GP, and $T_{\rm COBE}(z)$ corresponds to the standard TRR prediction with $T_0 = 2.72548~\text{K}$. 
The respective uncertainties are given by $\sigma_{\rm GP}(z)$ 
and $\sigma_{\rm COBE}(z)$.

\begin{figure}[htbp]
\centering
\includegraphics[scale=0.6]{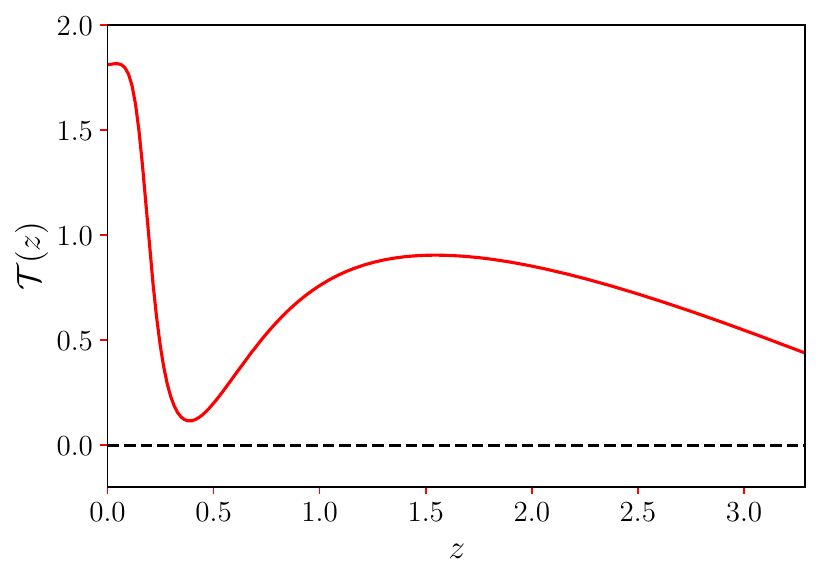} 
\caption{Evolution of the degree of tension between the statistical reconstructions developed in this work and the direct measurement from COBE/FIRAS, as a function of redshift \( z \).} 
\label{fig:tension}
\end{figure}

In Figure~\ref{fig:tension}, we present the degree of tension $\mathcal{T}(z)$ as a function of redshift. 
We observe deviations of less than $\sim$2$\sigma$ confidence level 
in the low-redshift regime, which gradually decrease until 
approximately $z \approx 0.5$. 
Beyond this point, the tension remains below 1$\sigma$ at higher redshifts. 
This behavior suggests a general agreement between the reconstructed temperature and the standard TRR for $z \gtrsim 0.5$, where the observed deviations can likely be attributed to statistical fluctuations within the uncertainties. 
For $z < 0.5$, and especially as $z \to 0$, the tension reaches up to $\sim$2$\sigma$, indicating a potential deviation that merits further investigation. 
This could point to either unaccounted systematics or a genuine departure from the standard cosmological expectations at late times.

In the next section, we discuss possible phenomenological implications of the results obtained in this section.


\section{Some Implications for phenomenological models}
\label{sec:Results2}
Certain classes of modified gravity theories that explicitly violate the Einstein Equivalence Principle (EEP) can be described by a matter action of the form (see \cite{2014PhRvD..90l4064H} and references therein):
\begin{equation}
    S_m = \sum_i \int d^4x\, \sqrt{-g}\, h_i(\phi)\, \mathcal{L}_i(g_{\mu\nu}, \Psi_i),
\end{equation}
where \( \mathcal{L}_i \) denotes the Lagrangian density of each matter field \( \Psi_i \), and \( h_i(\phi) \) encodes a non-minimal coupling between these fields and a scalar field \( \phi \). This coupling is a common feature in various alternative theories of gravity. When \( h_i(\phi) = 1 \), the standard framework of general relativity (GR) is recovered.

In particular, if the electromagnetic sector couples to \( \phi \), the fine-structure constant \( \alpha \) becomes a dynamical quantity. In this scenario, one finds \( \alpha \propto h^{-1}(\phi(t)) \), implying that the electromagnetic interaction strength may vary with cosmic time \cite{2014PhRvD..90l4064H}. As a result, the entire electromagnetic sector is affected, giving rise to observable consequences such as deviations from the CDDR, written as \( D_L (1+z)^2 / D_A = \eta(z) \), variations in the fine-structure constant \( \Delta\alpha/\alpha \), and changes in the redshift evolution law of the CMB temperature.

Within this framework, it has been shown that variations in the fine-structure constant and deviations in the CMB temperature evolution are intimately and unequivocally linked through the relation \cite{2014PhRvD..90l4064H}:
\begin{equation}
\label{alpha_th}
\frac{\Delta\alpha(z)}{\alpha} + 1 = 8.33 \, \frac{T_{\rm CMB}(z)}{T_0 (1+z)^{-7.33}} \,,
\end{equation}
where \( \alpha \) is the fine-structure constant and \( T_{\rm CMB}(z) \) is the CMB temperature at redshift \( z \). This expression highlights how modifications in the electromagnetic sector induced by the scalar field \( \phi \) can be simultaneously constrained by multiple independent cosmological probes~\cite{2024EPJC...84.1120F}.

\begin{figure}[htbp]
\centering
\includegraphics[scale=0.6]{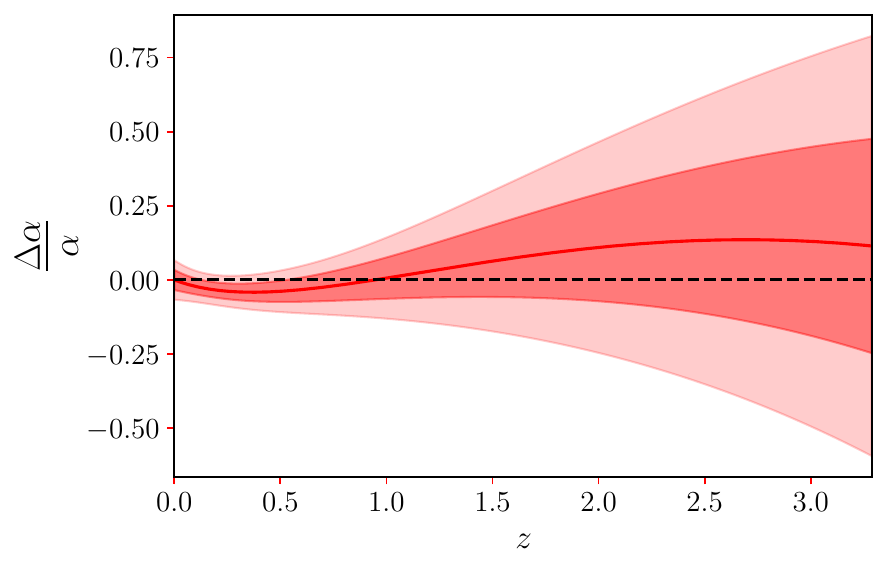}
\caption{GP reconstruction of the fine-structure constant using the SE kernel. The shaded regions represent the 1$\sigma$ and 2$\sigma$ statistical confidence intervals. The black dashed line correspond to the expect value.}
\label{fig:alfa_recon}
\end{figure}

In Figure~\ref{fig:alfa_recon}, we present the statistical reconstruction of the relative variation of the fine-structure constant, \( \Delta\alpha(z)/\alpha \), following the same GP methodology outlined in the previous sections. 
This reconstruction is obtained by applying equation~(\ref{alpha_th}), which connects the temperature evolution of the CMB to variations in the electromagnetic sector under the class of modified gravity theories discussed. 
Our results indicate that the constraints on \( \Delta\alpha(z)/\alpha \) are more stringent at low redshifts, particularly for \( z < 0.5 \), where the number of measurements is greater and more accurate. 
As redshift increases, the associated uncertainties grow, reflecting the decreasing sensitivity of current data at higher \( z \). Importantly, we do not observe statistically significant deviations from the null hypothesis \( \Delta\alpha(z)/\alpha = 0 \). This outcome suggests full consistency with the predictions of GR and the standard cosmological model. 
It is interesting to note that the redshift evolution of \( \Delta\alpha(z)/\alpha \) approximately follows the same shape as that of $\beta(z)$. This suggests that a variation of the fine-structure constant could also be a possible interpretation for our results regarding $\beta$. It is important to highlight that the analyses presented in this work are non-parametric and, therefore, minimally dependent on specific physical --or cosmological-- assumptions. As such, a more detailed investigation aimed at explaining the observed shape of $\beta(z)$, based on particular theoretical models, may serve as a motivation for future studies.

It is worth emphasizing that although our reconstruction of \( \beta(z) \) in the previous section showed hints of possible deviations at low redshifts, such effect do not automatically translates into evidence for new physics in other related observables. In particular, the lack of significant variation in \( \alpha \) highlights the importance of a multi-probe approach: consistency across different sectors strengthens the robustness of our results and places tighter constraints on theoretical models that predict coupled variations in fundamental constants.

\section{Final Remarks}
\label{final}

In this work, we performed a model-independent statistical reconstruction of the CMB temperature evolution, \( T_{\rm CMB}(z) \), using GP techniques applied to two observational datasets: measurements from the Sunyaev-Zel’dovich effect and data related to CMB–Interstellar Medium interactions. 
Our analyses revealed possible deviations from the standard TRR, \( T_{\rm CMB}(z) = T_0 (1+z) \), particularly at low redshifts (\( z < 0.5 \)), where the reconstructed curve mildly departs from the $\Lambda$CDM prediction at the \(\sim 2\sigma\) level. We also identified a mild tension between our inference of \( T_0 \) and the value measured by COBE/FIRAS, which may be indicative of local deviations or just statistical fluctuations.

To better understand the potential theoretical implications of these results, we explored their connection with phenomenological models that predict a coupling between the electromagnetic sector and a scalar field, leading to violations of the EEP. Within this framework, we derived a reconstruction for the relative variation of the fine-structure constant, \( \Delta\alpha(z)/\alpha \), based on the inferred CMB temperature evolution. Our findings show no significant evidence of a time variation in \( \alpha \), with results consistent with the null hypothesis and standard physics expectations across the entire redshift range analyzed. 

These results highlight the importance of adopting multi-probe approaches when testing the foundations of cosmology. While small deviations in temperature evolution may suggest the need for further investigation, the lack of correlated variations in the fine-structure constant reinforces the robustness of standard cosmological assumptions. Future improvements in low-redshift measurements and independent constraints on \( \alpha \) will be crucial to better clarify whether the observed deviations are purely statistical or hint a new physics beyond GR.

\begin{acknowledgments}
  
FA thanks to Funda\c{c}\~{a}o Carlos Chagas Filho de Amparo \`{a} Pesquisa do Estado do Rio de Janeiro (FAPERJ), Processo SEI-260003/001221/2025, for the financial support.  RFLH thanks to CNPq support under the project No.308550/2023-4.
ABR is supported by FAPERJ, Grant No E-26/200.149/2025 and 200.150/2025 (304809). ABR would also like to thank CERN-TH for supporting the "short visit" program and for providing its facilities, where part of this work was carried out. RCN thanks the financial support from CNPq under the project No. 304306/2022-3 and FAPERGS for partial financial support under the project No. 23/2551-0000848-3. 
AB acknowledges a CNPq fellowship. 
This article is based on work from COST Action CA21136 - “Addressing observational tensions in cosmology with systematics and fundamental physics (CosmoVerse)", supported by COST (European Cooperation in Science and Technology). 


\end{acknowledgments}

\bibliographystyle{apsrev4-1}
\bibliography{main}

\begin{thebibliography}{61}%
\makeatletter
\providecommand \@ifxundefined [1]{%
 \@ifx{#1\undefined}
}%
\providecommand \@ifnum [1]{%
 \ifnum #1\expandafter \@firstoftwo
 \else \expandafter \@secondoftwo
 \fi
}%
\providecommand \@ifx [1]{%
 \ifx #1\expandafter \@firstoftwo
 \else \expandafter \@secondoftwo
 \fi
}%
\providecommand \natexlab [1]{#1}%
\providecommand \enquote  [1]{``#1''}%
\providecommand \bibnamefont  [1]{#1}%
\providecommand \bibfnamefont [1]{#1}%
\providecommand \citenamefont [1]{#1}%
\providecommand \href@noop [0]{\@secondoftwo}%
\providecommand \href [0]{\begingroup \@sanitize@url \@href}%
\providecommand \@href[1]{\@@startlink{#1}\@@href}%
\providecommand \@@href[1]{\endgroup#1\@@endlink}%
\providecommand \@sanitize@url [0]{\catcode `\\12\catcode `\$12\catcode
  `\&12\catcode `\#12\catcode `\^12\catcode `\_12\catcode `\%12\relax}%
\providecommand \@@startlink[1]{}%
\providecommand \@@endlink[0]{}%
\providecommand \url  [0]{\begingroup\@sanitize@url \@url }%
\providecommand \@url [1]{\endgroup\@href {#1}{\urlprefix }}%
\providecommand \urlprefix  [0]{URL }%
\providecommand \Eprint [0]{\href }%
\providecommand \doibase [0]{http://dx.doi.org/}%
\providecommand \selectlanguage [0]{\@gobble}%
\providecommand \bibinfo  [0]{\@secondoftwo}%
\providecommand \bibfield  [0]{\@secondoftwo}%
\providecommand \translation [1]{[#1]}%
\providecommand \BibitemOpen [0]{}%
\providecommand \bibitemStop [0]{}%
\providecommand \bibitemNoStop [0]{.\EOS\space}%
\providecommand \EOS [0]{\spacefactor3000\relax}%
\providecommand \BibitemShut  [1]{\csname bibitem#1\endcsname}%
\let\auto@bib@innerbib\@empty
\bibitem [{\citenamefont {{Hu}}\ and\ \citenamefont {{Dodelson}}(2002)}]{Hu02}%
  \BibitemOpen
  \bibfield  {author} {\bibinfo {author} {\bibfnamefont {W.}~\bibnamefont
  {{Hu}}}\ and\ \bibinfo {author} {\bibfnamefont {S.}~\bibnamefont
  {{Dodelson}}},\ }\href {\doibase 10.1146/annurev.astro.40.060401.093926}
  {\bibfield  {journal} {\bibinfo  {journal} {araa}\ }\textbf {\bibinfo
  {volume} {40}},\ \bibinfo {pages} {171} (\bibinfo {year} {2002})},\ \Eprint
  {http://arxiv.org/abs/astro-ph/0110414} {arXiv:astro-ph/0110414 [astro-ph]}
  \BibitemShut {NoStop}%
\bibitem [{\citenamefont {Dodelson}\ and\ \citenamefont
  {Schmidt}(2024)}]{Dodelson_book}%
  \BibitemOpen
  \bibfield  {author} {\bibinfo {author} {\bibfnamefont {S.}~\bibnamefont
  {Dodelson}}\ and\ \bibinfo {author} {\bibfnamefont {F.}~\bibnamefont
  {Schmidt}},\ }\href@noop {} {\emph {\bibinfo {title} {Modern cosmology}}}\
  (\bibinfo  {publisher} {Elsevier},\ \bibinfo {year} {2024})\BibitemShut
  {NoStop}%
\bibitem [{\citenamefont {Collaboration}(2020)}]{Planck20}%
  \BibitemOpen
  \bibfield  {author} {\bibinfo {author} {\bibfnamefont {P.}~\bibnamefont
  {Collaboration}},\ }\href {\doibase 10.1051/0004-6361/201833910} {\bibfield
  {journal} {\bibinfo  {journal} {aap}\ }\textbf {\bibinfo {volume} {641}},\
  \bibinfo {eid} {A6} (\bibinfo {year} {2020})},\ \Eprint
  {http://arxiv.org/abs/1807.06209} {arXiv:1807.06209 [astro-ph.CO]}
  \BibitemShut {NoStop}%
\bibitem [{\citenamefont {Louis}\ \emph {et~al.}(2025)\citenamefont {Louis}
  \emph {et~al.}}]{ACT:2025fju}%
  \BibitemOpen
  \bibfield  {author} {\bibinfo {author} {\bibfnamefont {T.}~\bibnamefont
  {Louis}} \emph {et~al.} (\bibinfo {collaboration} {ACT}),\ }\href@noop {}
  {\bibfield  {journal} {\bibinfo  {journal} {arXiv e-prints}\ } (\bibinfo
  {year} {2025})},\ \Eprint {http://arxiv.org/abs/2503.14452} {arXiv:2503.14452
  [astro-ph.CO]} \BibitemShut {NoStop}%
\bibitem [{\citenamefont {Saro}\ \emph {et~al.}(2014)\citenamefont {Saro} \emph
  {et~al.}}]{Saro14}%
  \BibitemOpen
  \bibfield  {author} {\bibinfo {author} {\bibfnamefont {A.}~\bibnamefont
  {Saro}} \emph {et~al.} (\bibinfo {collaboration} {SPT}),\ }\href {\doibase
  10.1093/mnras/stu575} {\bibfield  {journal} {\bibinfo  {journal} {Mon. Not.
  Roy. Astron. Soc.}\ }\textbf {\bibinfo {volume} {440}},\ \bibinfo {pages}
  {2610} (\bibinfo {year} {2014})},\ \Eprint {http://arxiv.org/abs/1312.2462}
  {arXiv:1312.2462 [astro-ph.CO]} \BibitemShut {NoStop}%
\bibitem [{\citenamefont {{Bernui}}\ and\ \citenamefont
  {{Rebou{\c{c}}as}}(2010)}]{Bernui2010}%
  \BibitemOpen
  \bibfield  {author} {\bibinfo {author} {\bibfnamefont {A.}~\bibnamefont
  {{Bernui}}}\ and\ \bibinfo {author} {\bibfnamefont {M.~J.}\ \bibnamefont
  {{Rebou{\c{c}}as}}},\ }\href {\doibase 10.1103/PhysRevD.81.063533} {\bibfield
   {journal} {\bibinfo  {journal} {prd}\ }\textbf {\bibinfo {volume} {81}},\
  \bibinfo {eid} {063533} (\bibinfo {year} {2010})},\ \Eprint
  {http://arxiv.org/abs/0912.0269} {arXiv:0912.0269 [astro-ph.CO]} \BibitemShut
  {NoStop}%
\bibitem [{\citenamefont {{Bernui}}\ and\ \citenamefont
  {{Rebou{\c{c}}as}}(2009)}]{Bernui2009}%
  \BibitemOpen
  \bibfield  {author} {\bibinfo {author} {\bibfnamefont {A.}~\bibnamefont
  {{Bernui}}}\ and\ \bibinfo {author} {\bibfnamefont {M.~J.}\ \bibnamefont
  {{Rebou{\c{c}}as}}},\ }\href {\doibase 10.1103/PhysRevD.79.063528} {\bibfield
   {journal} {\bibinfo  {journal} {prd}\ }\textbf {\bibinfo {volume} {79}},\
  \bibinfo {eid} {063528} (\bibinfo {year} {2009})},\ \Eprint
  {http://arxiv.org/abs/0806.3758} {arXiv:0806.3758 [astro-ph]} \BibitemShut
  {NoStop}%
\bibitem [{\citenamefont {{Bernui}}\ \emph {et~al.}(2007)\citenamefont
  {{Bernui}}, \citenamefont {{Mota}}, \citenamefont {{Rebou{\c{c}}as}},\ and\
  \citenamefont {{Tavakol}}}]{BMRT2007IJMPD}%
  \BibitemOpen
  \bibfield  {author} {\bibinfo {author} {\bibfnamefont {A.}~\bibnamefont
  {{Bernui}}}, \bibinfo {author} {\bibfnamefont {B.}~\bibnamefont {{Mota}}},
  \bibinfo {author} {\bibfnamefont {M.~J.}\ \bibnamefont {{Rebou{\c{c}}as}}}, \
  and\ \bibinfo {author} {\bibfnamefont {R.}~\bibnamefont {{Tavakol}}},\ }\href
  {\doibase 10.1142/S0218271807010195} {\bibfield  {journal} {\bibinfo
  {journal} {International Journal of Modern Physics D}\ }\textbf {\bibinfo
  {volume} {16}},\ \bibinfo {pages} {411} (\bibinfo {year} {2007})},\ \Eprint
  {http://arxiv.org/abs/0706.0575} {arXiv:0706.0575 [astro-ph]} \BibitemShut
  {NoStop}%
\bibitem [{\citenamefont {{Gruppuso}}(2014)}]{Gruppuso2014}%
  \BibitemOpen
  \bibfield  {author} {\bibinfo {author} {\bibfnamefont {A.}~\bibnamefont
  {{Gruppuso}}},\ }\href {\doibase 10.1093/mnras/stt1937} {\bibfield  {journal}
  {\bibinfo  {journal} {mnras}\ }\textbf {\bibinfo {volume} {437}},\ \bibinfo
  {pages} {2076} (\bibinfo {year} {2014})},\ \Eprint
  {http://arxiv.org/abs/1310.2822} {arXiv:1310.2822 [astro-ph.CO]} \BibitemShut
  {NoStop}%
\bibitem [{\citenamefont {{Planck Collaboration}}(2020)}]{Planck-iso+stat2020}%
  \BibitemOpen
  \bibfield  {author} {\bibinfo {author} {\bibnamefont {{Planck
  Collaboration}}},\ }\href {\doibase 10.1051/0004-6361/201935201} {\bibfield
  {journal} {\bibinfo  {journal} {aap}\ }\textbf {\bibinfo {volume} {641}},\
  \bibinfo {eid} {A7} (\bibinfo {year} {2020})},\ \Eprint
  {http://arxiv.org/abs/1906.02552} {arXiv:1906.02552 [astro-ph.CO]}
  \BibitemShut {NoStop}%
\bibitem [{\citenamefont {{Smoot}}\ \emph {et~al.}(1987)\citenamefont
  {{Smoot}}, \citenamefont {{Bensadoun}}, \citenamefont {{Bersanelli}},
  \citenamefont {{de Amici}}, \citenamefont {{Kogut}}, \citenamefont
  {{Levin}},\ and\ \citenamefont {{Witebsky}}}]{Smoot1987}%
  \BibitemOpen
  \bibfield  {author} {\bibinfo {author} {\bibfnamefont {G.~F.}\ \bibnamefont
  {{Smoot}}}, \bibinfo {author} {\bibfnamefont {M.}~\bibnamefont
  {{Bensadoun}}}, \bibinfo {author} {\bibfnamefont {M.}~\bibnamefont
  {{Bersanelli}}}, \bibinfo {author} {\bibfnamefont {G.}~\bibnamefont {{de
  Amici}}}, \bibinfo {author} {\bibfnamefont {A.}~\bibnamefont {{Kogut}}},
  \bibinfo {author} {\bibfnamefont {S.}~\bibnamefont {{Levin}}}, \ and\
  \bibinfo {author} {\bibfnamefont {C.}~\bibnamefont {{Witebsky}}},\ }\href
  {\doibase 10.1086/184910} {\bibfield  {journal} {\bibinfo  {journal} {apjl}\
  }\textbf {\bibinfo {volume} {317}},\ \bibinfo {pages} {L45} (\bibinfo {year}
  {1987})}\BibitemShut {NoStop}%
\bibitem [{\citenamefont {{Mather}}\ \emph {et~al.}(1994)\citenamefont
  {{Mather}}, \citenamefont {{Cheng}}, \citenamefont {{Cottingham}},
  \citenamefont {{Eplee}}, \citenamefont {{Fixsen}}, \citenamefont
  {{Hewagama}}, \citenamefont {{Isaacman}}, \citenamefont {{Jensen}},
  \citenamefont {{Meyer}}, \citenamefont {{Noerdlinger}}, \citenamefont
  {{Read}}, \citenamefont {{Rosen}}, \citenamefont {{Shafer}}, \citenamefont
  {{Wright}}, \citenamefont {{Bennett}}, \citenamefont {{Boggess}},
  \citenamefont {{Hauser}}, \citenamefont {{Kelsall}}, \citenamefont
  {{Moseley}}, \citenamefont {{Silverberg}}, \citenamefont {{Smoot}},
  \citenamefont {{Weiss}},\ and\ \citenamefont {{Wilkinson}}}]{Mather94}%
  \BibitemOpen
  \bibfield  {author} {\bibinfo {author} {\bibfnamefont {J.~C.}\ \bibnamefont
  {{Mather}}}, \bibinfo {author} {\bibfnamefont {E.~S.}\ \bibnamefont
  {{Cheng}}}, \bibinfo {author} {\bibfnamefont {D.~A.}\ \bibnamefont
  {{Cottingham}}}, \bibinfo {author} {\bibfnamefont {R.~E.}\ \bibnamefont
  {{Eplee}}, \bibfnamefont {Jr.}}, \bibinfo {author} {\bibfnamefont {D.~J.}\
  \bibnamefont {{Fixsen}}}, \bibinfo {author} {\bibfnamefont {T.}~\bibnamefont
  {{Hewagama}}}, \bibinfo {author} {\bibfnamefont {R.~B.}\ \bibnamefont
  {{Isaacman}}}, \bibinfo {author} {\bibfnamefont {K.~A.}\ \bibnamefont
  {{Jensen}}}, \bibinfo {author} {\bibfnamefont {S.~S.}\ \bibnamefont
  {{Meyer}}}, \bibinfo {author} {\bibfnamefont {P.~D.}\ \bibnamefont
  {{Noerdlinger}}}, \bibinfo {author} {\bibfnamefont {S.~M.}\ \bibnamefont
  {{Read}}}, \bibinfo {author} {\bibfnamefont {L.~P.}\ \bibnamefont {{Rosen}}},
  \bibinfo {author} {\bibfnamefont {R.~A.}\ \bibnamefont {{Shafer}}}, \bibinfo
  {author} {\bibfnamefont {E.~L.}\ \bibnamefont {{Wright}}}, \bibinfo {author}
  {\bibfnamefont {C.~L.}\ \bibnamefont {{Bennett}}}, \bibinfo {author}
  {\bibfnamefont {N.~W.}\ \bibnamefont {{Boggess}}}, \bibinfo {author}
  {\bibfnamefont {M.~G.}\ \bibnamefont {{Hauser}}}, \bibinfo {author}
  {\bibfnamefont {T.}~\bibnamefont {{Kelsall}}}, \bibinfo {author}
  {\bibfnamefont {S.~H.}\ \bibnamefont {{Moseley}}, \bibfnamefont {Jr.}},
  \bibinfo {author} {\bibfnamefont {R.~F.}\ \bibnamefont {{Silverberg}}},
  \bibinfo {author} {\bibfnamefont {G.~F.}\ \bibnamefont {{Smoot}}}, \bibinfo
  {author} {\bibfnamefont {R.}~\bibnamefont {{Weiss}}}, \ and\ \bibinfo
  {author} {\bibfnamefont {D.~T.}\ \bibnamefont {{Wilkinson}}},\ }\href
  {\doibase 10.1086/173574} {\bibfield  {journal} {\bibinfo  {journal} {apj}\
  }\textbf {\bibinfo {volume} {420}},\ \bibinfo {pages} {439} (\bibinfo {year}
  {1994})}\BibitemShut {NoStop}%
\bibitem [{\citenamefont {{Fixsen}}\ \emph {et~al.}(1996)\citenamefont
  {{Fixsen}}, \citenamefont {{Cheng}}, \citenamefont {{Gales}}, \citenamefont
  {{Mather}}, \citenamefont {{Shafer}},\ and\ \citenamefont
  {{Wright}}}]{Fixen96}%
  \BibitemOpen
  \bibfield  {author} {\bibinfo {author} {\bibfnamefont {D.~J.}\ \bibnamefont
  {{Fixsen}}}, \bibinfo {author} {\bibfnamefont {E.~S.}\ \bibnamefont
  {{Cheng}}}, \bibinfo {author} {\bibfnamefont {J.~M.}\ \bibnamefont
  {{Gales}}}, \bibinfo {author} {\bibfnamefont {J.~C.}\ \bibnamefont
  {{Mather}}}, \bibinfo {author} {\bibfnamefont {R.~A.}\ \bibnamefont
  {{Shafer}}}, \ and\ \bibinfo {author} {\bibfnamefont {E.~L.}\ \bibnamefont
  {{Wright}}},\ }\href {\doibase 10.1086/178173} {\bibfield  {journal}
  {\bibinfo  {journal} {apj}\ }\textbf {\bibinfo {volume} {473}},\ \bibinfo
  {pages} {576} (\bibinfo {year} {1996})},\ \Eprint
  {http://arxiv.org/abs/astro-ph/9605054} {arXiv:astro-ph/9605054 [astro-ph]}
  \BibitemShut {NoStop}%
\bibitem [{\citenamefont {{Klimenko}}\ \emph {et~al.}(2020)\citenamefont
  {{Klimenko}}, \citenamefont {{Ivanchik}}, \citenamefont {{Petitjean}},
  \citenamefont {{Noterdaeme}},\ and\ \citenamefont {{Srianand}}}]{Klimenko20}%
  \BibitemOpen
  \bibfield  {author} {\bibinfo {author} {\bibfnamefont {V.~V.}\ \bibnamefont
  {{Klimenko}}}, \bibinfo {author} {\bibfnamefont {A.~V.}\ \bibnamefont
  {{Ivanchik}}}, \bibinfo {author} {\bibfnamefont {P.}~\bibnamefont
  {{Petitjean}}}, \bibinfo {author} {\bibfnamefont {P.}~\bibnamefont
  {{Noterdaeme}}}, \ and\ \bibinfo {author} {\bibfnamefont {R.}~\bibnamefont
  {{Srianand}}},\ }\href {\doibase 10.1134/S1063773720110031} {\bibfield
  {journal} {\bibinfo  {journal} {Astronomy Letters}\ }\textbf {\bibinfo
  {volume} {46}},\ \bibinfo {pages} {715} (\bibinfo {year} {2020})},\ \Eprint
  {http://arxiv.org/abs/2106.00119} {arXiv:2106.00119 [astro-ph.CO]}
  \BibitemShut {NoStop}%
\bibitem [{\citenamefont {{Bersanelli}}\ \emph {et~al.}(2002)\citenamefont
  {{Bersanelli}}, \citenamefont {{Maino}},\ and\ \citenamefont
  {{Mennella}}}]{Bersanelli2002}%
  \BibitemOpen
  \bibfield  {author} {\bibinfo {author} {\bibfnamefont {M.}~\bibnamefont
  {{Bersanelli}}}, \bibinfo {author} {\bibfnamefont {D.}~\bibnamefont
  {{Maino}}}, \ and\ \bibinfo {author} {\bibfnamefont {A.}~\bibnamefont
  {{Mennella}}},\ }\href {\doibase 10.48550/arXiv.astro-ph/0209215} {\bibfield
  {journal} {\bibinfo  {journal} {Nuovo Cimento Rivista Serie}\ }\textbf
  {\bibinfo {volume} {025}},\ \bibinfo {pages} {1} (\bibinfo {year} {2002})},\
  \Eprint {http://arxiv.org/abs/astro-ph/0209215} {arXiv:astro-ph/0209215
  [astro-ph]} \BibitemShut {NoStop}%
\bibitem [{\citenamefont {{Fixsen}}(2009)}]{Fixen09}%
  \BibitemOpen
  \bibfield  {author} {\bibinfo {author} {\bibfnamefont {D.~J.}\ \bibnamefont
  {{Fixsen}}},\ }\href {\doibase 10.1088/0004-637X/707/2/916} {\bibfield
  {journal} {\bibinfo  {journal} {apj}\ }\textbf {\bibinfo {volume} {707}},\
  \bibinfo {pages} {916} (\bibinfo {year} {2009})},\ \Eprint
  {http://arxiv.org/abs/0911.1955} {arXiv:0911.1955 [astro-ph.CO]} \BibitemShut
  {NoStop}%
\bibitem [{\citenamefont {Ivanov}\ \emph {et~al.}(2020)\citenamefont {Ivanov},
  \citenamefont {Ali-Haïmoud},\ and\ \citenamefont {Lesgourgues}}]{Ivanov20}%
  \BibitemOpen
  \bibfield  {author} {\bibinfo {author} {\bibfnamefont {M.~M.}\ \bibnamefont
  {Ivanov}}, \bibinfo {author} {\bibfnamefont {Y.}~\bibnamefont
  {Ali-Haïmoud}}, \ and\ \bibinfo {author} {\bibfnamefont {J.}~\bibnamefont
  {Lesgourgues}},\ }\href {\doibase 10.1103/physrevd.102.063515} {\bibfield
  {journal} {\bibinfo  {journal} {Physical Review D}\ }\textbf {\bibinfo
  {volume} {102}} (\bibinfo {year} {2020}),\
  10.1103/physrevd.102.063515}\BibitemShut {NoStop}%
\bibitem [{\citenamefont {Gush}\ \emph {et~al.}(1990)\citenamefont {Gush},
  \citenamefont {Halpern},\ and\ \citenamefont {Wishnow}}]{Gush90}%
  \BibitemOpen
  \bibfield  {author} {\bibinfo {author} {\bibfnamefont {H.~P.}\ \bibnamefont
  {Gush}}, \bibinfo {author} {\bibfnamefont {M.}~\bibnamefont {Halpern}}, \
  and\ \bibinfo {author} {\bibfnamefont {E.~H.}\ \bibnamefont {Wishnow}},\
  }\href {\doibase 10.1103/PhysRevLett.65.537} {\bibfield  {journal} {\bibinfo
  {journal} {Phys. Rev. Lett.}\ }\textbf {\bibinfo {volume} {65}},\ \bibinfo
  {pages} {537} (\bibinfo {year} {1990})}\BibitemShut {NoStop}%
\bibitem [{\citenamefont {{Noterdaeme}}\ \emph {et~al.}(2011)\citenamefont
  {{Noterdaeme}}, \citenamefont {{Petitjean}}, \citenamefont {{Srianand}},
  \citenamefont {{Ledoux}},\ and\ \citenamefont {{L{\'o}pez}}}]{Noterdaeme11}%
  \BibitemOpen
  \bibfield  {author} {\bibinfo {author} {\bibfnamefont {P.}~\bibnamefont
  {{Noterdaeme}}}, \bibinfo {author} {\bibfnamefont {P.}~\bibnamefont
  {{Petitjean}}}, \bibinfo {author} {\bibfnamefont {R.}~\bibnamefont
  {{Srianand}}}, \bibinfo {author} {\bibfnamefont {C.}~\bibnamefont
  {{Ledoux}}}, \ and\ \bibinfo {author} {\bibfnamefont {S.}~\bibnamefont
  {{L{\'o}pez}}},\ }\href {\doibase 10.1051/0004-6361/201016140} {\bibfield
  {journal} {\bibinfo  {journal} {aap}\ }\textbf {\bibinfo {volume} {526}},\
  \bibinfo {eid} {L7} (\bibinfo {year} {2011})},\ \Eprint
  {http://arxiv.org/abs/1012.3164} {arXiv:1012.3164 [astro-ph.CO]} \BibitemShut
  {NoStop}%
\bibitem [{\citenamefont {{Guo}}\ and\ \citenamefont {{Zhang}}(2016)}]{Guo16}%
  \BibitemOpen
  \bibfield  {author} {\bibinfo {author} {\bibfnamefont {R.-Y.}\ \bibnamefont
  {{Guo}}}\ and\ \bibinfo {author} {\bibfnamefont {X.}~\bibnamefont
  {{Zhang}}},\ }\href {\doibase 10.1140/epjc/s10052-016-4016-x} {\bibfield
  {journal} {\bibinfo  {journal} {European Physical Journal C}\ }\textbf
  {\bibinfo {volume} {76}},\ \bibinfo {eid} {163} (\bibinfo {year} {2016})},\
  \Eprint {http://arxiv.org/abs/1512.07703} {arXiv:1512.07703 [astro-ph.CO]}
  \BibitemShut {NoStop}%
\bibitem [{\citenamefont {{Abitbol}}\ \emph {et~al.}(2020)\citenamefont
  {{Abitbol}}, \citenamefont {{Hill}},\ and\ \citenamefont
  {{Chluba}}}]{Abitbol20}%
  \BibitemOpen
  \bibfield  {author} {\bibinfo {author} {\bibfnamefont {M.~H.}\ \bibnamefont
  {{Abitbol}}}, \bibinfo {author} {\bibfnamefont {J.~C.}\ \bibnamefont
  {{Hill}}}, \ and\ \bibinfo {author} {\bibfnamefont {J.}~\bibnamefont
  {{Chluba}}},\ }\href {\doibase 10.3847/1538-4357/ab7b70} {\bibfield
  {journal} {\bibinfo  {journal} {apj}\ }\textbf {\bibinfo {volume} {893}},\
  \bibinfo {eid} {18} (\bibinfo {year} {2020})},\ \Eprint
  {http://arxiv.org/abs/1910.09881} {arXiv:1910.09881 [astro-ph.CO]}
  \BibitemShut {NoStop}%
\bibitem [{\citenamefont {{Luzzi}}\ \emph {et~al.}(2009)\citenamefont
  {{Luzzi}}, \citenamefont {{Shimon}}, \citenamefont {{Lamagna}}, \citenamefont
  {{Rephaeli}}, \citenamefont {{De Petris}}, \citenamefont {{Conte}},
  \citenamefont {{De Gregori}},\ and\ \citenamefont
  {{Battistelli}}}]{Luzzi2009}%
  \BibitemOpen
  \bibfield  {author} {\bibinfo {author} {\bibfnamefont {G.}~\bibnamefont
  {{Luzzi}}}, \bibinfo {author} {\bibfnamefont {M.}~\bibnamefont {{Shimon}}},
  \bibinfo {author} {\bibfnamefont {L.}~\bibnamefont {{Lamagna}}}, \bibinfo
  {author} {\bibfnamefont {Y.}~\bibnamefont {{Rephaeli}}}, \bibinfo {author}
  {\bibfnamefont {M.}~\bibnamefont {{De Petris}}}, \bibinfo {author}
  {\bibfnamefont {A.}~\bibnamefont {{Conte}}}, \bibinfo {author} {\bibfnamefont
  {S.}~\bibnamefont {{De Gregori}}}, \ and\ \bibinfo {author} {\bibfnamefont
  {E.~S.}\ \bibnamefont {{Battistelli}}},\ }\href {\doibase
  10.1088/0004-637X/705/2/1122} {\bibfield  {journal} {\bibinfo  {journal}
  {apj}\ }\textbf {\bibinfo {volume} {705}},\ \bibinfo {pages} {1122} (\bibinfo
  {year} {2009})},\ \Eprint {http://arxiv.org/abs/0909.2815} {arXiv:0909.2815
  [astro-ph.CO]} \BibitemShut {NoStop}%
\bibitem [{\citenamefont {{Luzzi}}\ \emph {et~al.}(2015)\citenamefont
  {{Luzzi}}, \citenamefont {{G{\'e}nova-Santos}}, \citenamefont {{Martins}},
  \citenamefont {{De Petris}},\ and\ \citenamefont {{Lamagna}}}]{Luzzi15}%
  \BibitemOpen
  \bibfield  {author} {\bibinfo {author} {\bibfnamefont {G.}~\bibnamefont
  {{Luzzi}}}, \bibinfo {author} {\bibfnamefont {R.~T.}\ \bibnamefont
  {{G{\'e}nova-Santos}}}, \bibinfo {author} {\bibfnamefont {C.~J.~A.~P.}\
  \bibnamefont {{Martins}}}, \bibinfo {author} {\bibfnamefont {M.}~\bibnamefont
  {{De Petris}}}, \ and\ \bibinfo {author} {\bibfnamefont {L.}~\bibnamefont
  {{Lamagna}}},\ }\href {\doibase 10.1088/1475-7516/2015/09/011} {\bibfield
  {journal} {\bibinfo  {journal} {jcap}\ }\textbf {\bibinfo {volume} {2015}},\
  \bibinfo {pages} {011} (\bibinfo {year} {2015})},\ \Eprint
  {http://arxiv.org/abs/1502.07858} {arXiv:1502.07858 [astro-ph.CO]}
  \BibitemShut {NoStop}%
\bibitem [{\citenamefont {{de Martino}}\ \emph {et~al.}(2015)\citenamefont {{de
  Martino}}, \citenamefont {{G{\'e}nova-Santos}}, \citenamefont
  {{Atrio-Barandela}}, \citenamefont {{Ebeling}}, \citenamefont {{Kashlinsky}},
  \citenamefont {{Kocevski}},\ and\ \citenamefont {{Martins}}}]{Martino15}%
  \BibitemOpen
  \bibfield  {author} {\bibinfo {author} {\bibfnamefont {I.}~\bibnamefont {{de
  Martino}}}, \bibinfo {author} {\bibfnamefont {R.}~\bibnamefont
  {{G{\'e}nova-Santos}}}, \bibinfo {author} {\bibfnamefont {F.}~\bibnamefont
  {{Atrio-Barandela}}}, \bibinfo {author} {\bibfnamefont {H.}~\bibnamefont
  {{Ebeling}}}, \bibinfo {author} {\bibfnamefont {A.}~\bibnamefont
  {{Kashlinsky}}}, \bibinfo {author} {\bibfnamefont {D.}~\bibnamefont
  {{Kocevski}}}, \ and\ \bibinfo {author} {\bibfnamefont {C.~J.~A.~P.}\
  \bibnamefont {{Martins}}},\ }\href {\doibase 10.1088/0004-637X/808/2/128}
  {\bibfield  {journal} {\bibinfo  {journal} {apj}\ }\textbf {\bibinfo {volume}
  {808}},\ \bibinfo {eid} {128} (\bibinfo {year} {2015})},\ \Eprint
  {http://arxiv.org/abs/1502.06707} {arXiv:1502.06707 [astro-ph.CO]}
  \BibitemShut {NoStop}%
\bibitem [{\citenamefont {{Arjona}}(2020)}]{Arjona20}%
  \BibitemOpen
  \bibfield  {author} {\bibinfo {author} {\bibfnamefont {R.}~\bibnamefont
  {{Arjona}}},\ }\href {\doibase 10.1088/1475-7516/2020/08/009} {\bibfield
  {journal} {\bibinfo  {journal} {jcap}\ }\textbf {\bibinfo {volume} {2020}},\
  \bibinfo {eid} {009} (\bibinfo {year} {2020})},\ \Eprint
  {http://arxiv.org/abs/2002.12700} {arXiv:2002.12700 [astro-ph.CO]}
  \BibitemShut {NoStop}%
\bibitem [{\citenamefont {Li}\ \emph {et~al.}(2021)\citenamefont {Li} \emph
  {et~al.}}]{Li20}%
  \BibitemOpen
  \bibfield  {author} {\bibinfo {author} {\bibfnamefont {Y.}~\bibnamefont {Li}}
  \emph {et~al.},\ }\href {\doibase 10.3847/1538-4357/ac26b6} {\bibfield
  {journal} {\bibinfo  {journal} {Astrophys. J.}\ }\textbf {\bibinfo {volume}
  {922}},\ \bibinfo {pages} {136} (\bibinfo {year} {2021})},\ \Eprint
  {http://arxiv.org/abs/2106.12467} {arXiv:2106.12467 [astro-ph.CO]}
  \BibitemShut {NoStop}%
\bibitem [{\citenamefont {{Riechers}}\ \emph {et~al.}(2022)\citenamefont
  {{Riechers}}, \citenamefont {{Weiss}}, \citenamefont {{Walter}},
  \citenamefont {{Carilli}}, \citenamefont {{Cox}}, \citenamefont {{Decarli}},\
  and\ \citenamefont {{Neri}}}]{Riechers22}%
  \BibitemOpen
  \bibfield  {author} {\bibinfo {author} {\bibfnamefont {D.~A.}\ \bibnamefont
  {{Riechers}}}, \bibinfo {author} {\bibfnamefont {A.}~\bibnamefont {{Weiss}}},
  \bibinfo {author} {\bibfnamefont {F.}~\bibnamefont {{Walter}}}, \bibinfo
  {author} {\bibfnamefont {C.~L.}\ \bibnamefont {{Carilli}}}, \bibinfo {author}
  {\bibfnamefont {P.}~\bibnamefont {{Cox}}}, \bibinfo {author} {\bibfnamefont
  {R.}~\bibnamefont {{Decarli}}}, \ and\ \bibinfo {author} {\bibfnamefont
  {R.}~\bibnamefont {{Neri}}},\ }\href {\doibase 10.1038/s41586-021-04294-5}
  {\bibfield  {journal} {\bibinfo  {journal} {nat}\ }\textbf {\bibinfo {volume}
  {602}},\ \bibinfo {pages} {58} (\bibinfo {year} {2022})},\ \Eprint
  {http://arxiv.org/abs/2202.00693} {arXiv:2202.00693 [astro-ph.GA]}
  \BibitemShut {NoStop}%
\bibitem [{\citenamefont {{Lima}}\ \emph {et~al.}(2000)\citenamefont {{Lima}},
  \citenamefont {{Silva}},\ and\ \citenamefont {{Viegas}}}]{Lima00}%
  \BibitemOpen
  \bibfield  {author} {\bibinfo {author} {\bibfnamefont {J.~A.~S.}\
  \bibnamefont {{Lima}}}, \bibinfo {author} {\bibfnamefont {A.~I.}\
  \bibnamefont {{Silva}}}, \ and\ \bibinfo {author} {\bibfnamefont {S.~M.}\
  \bibnamefont {{Viegas}}},\ }\href {\doibase 10.1046/j.1365-8711.2000.03172.x}
  {\bibfield  {journal} {\bibinfo  {journal} {mnras}\ }\textbf {\bibinfo
  {volume} {312}},\ \bibinfo {pages} {747} (\bibinfo {year}
  {2000})}\BibitemShut {NoStop}%
\bibitem [{\citenamefont {{Hees}}\ \emph {et~al.}(2014)\citenamefont {{Hees}},
  \citenamefont {{Minazzoli}},\ and\ \citenamefont
  {{Larena}}}]{2014PhRvD..90l4064H}%
  \BibitemOpen
  \bibfield  {author} {\bibinfo {author} {\bibfnamefont {A.}~\bibnamefont
  {{Hees}}}, \bibinfo {author} {\bibfnamefont {O.}~\bibnamefont {{Minazzoli}}},
  \ and\ \bibinfo {author} {\bibfnamefont {J.}~\bibnamefont {{Larena}}},\
  }\href {\doibase 10.1103/PhysRevD.90.124064} {\bibfield  {journal} {\bibinfo
  {journal} {prd}\ }\textbf {\bibinfo {volume} {90}},\ \bibinfo {eid} {124064}
  (\bibinfo {year} {2014})},\ \Eprint {http://arxiv.org/abs/1406.6187}
  {arXiv:1406.6187 [astro-ph.CO]} \BibitemShut {NoStop}%
\bibitem [{\citenamefont {{Ferreira}}\ \emph {et~al.}(2024)\citenamefont
  {{Ferreira}}, \citenamefont {{Holanda}}, \citenamefont {{Gonzalez}},
  \citenamefont {{Cola{\c{c}}o}},\ and\ \citenamefont
  {{Nunes}}}]{2024EPJC...84.1120F}%
  \BibitemOpen
  \bibfield  {author} {\bibinfo {author} {\bibfnamefont {M.}~\bibnamefont
  {{Ferreira}}}, \bibinfo {author} {\bibfnamefont {R.~F.~L.}\ \bibnamefont
  {{Holanda}}}, \bibinfo {author} {\bibfnamefont {J.~E.}\ \bibnamefont
  {{Gonzalez}}}, \bibinfo {author} {\bibfnamefont {L.~R.}\ \bibnamefont
  {{Cola{\c{c}}o}}}, \ and\ \bibinfo {author} {\bibfnamefont {R.~C.}\
  \bibnamefont {{Nunes}}},\ }\href {\doibase 10.1140/epjc/s10052-024-13468-0}
  {\bibfield  {journal} {\bibinfo  {journal} {European Physical Journal C}\
  }\textbf {\bibinfo {volume} {84}},\ \bibinfo {eid} {1120} (\bibinfo {year}
  {2024})},\ \Eprint {http://arxiv.org/abs/2410.21542} {arXiv:2410.21542
  [astro-ph.CO]} \BibitemShut {NoStop}%
\bibitem [{\citenamefont {{Fabbri}}(1981)}]{Fabri78}%
  \BibitemOpen
  \bibfield  {author} {\bibinfo {author} {\bibfnamefont {R.}~\bibnamefont
  {{Fabbri}}},\ }\href {\doibase 10.1007/BF00649478} {\bibfield  {journal}
  {\bibinfo  {journal} {apss}\ }\textbf {\bibinfo {volume} {77}},\ \bibinfo
  {pages} {529} (\bibinfo {year} {1981})}\BibitemShut {NoStop}%
\bibitem [{\citenamefont {{Rephaeli}}(1980)}]{Rephaeli80}%
  \BibitemOpen
  \bibfield  {author} {\bibinfo {author} {\bibfnamefont {Y.}~\bibnamefont
  {{Rephaeli}}},\ }\href {\doibase 10.1086/158398} {\bibfield  {journal}
  {\bibinfo  {journal} {apj}\ }\textbf {\bibinfo {volume} {241}},\ \bibinfo
  {pages} {858} (\bibinfo {year} {1980})}\BibitemShut {NoStop}%
\bibitem [{\citenamefont {Battistelli}\ \emph {et~al.}(2002)\citenamefont
  {Battistelli}, \citenamefont {De~Petris}, \citenamefont {Lamagna},
  \citenamefont {Melchiorri}, \citenamefont {Palladino}, \citenamefont
  {Savini}, \citenamefont {Cooray}, \citenamefont {Melchiorri}, \citenamefont
  {Rephaeli},\ and\ \citenamefont {Shimon}}]{Battistelli_2002}%
  \BibitemOpen
  \bibfield  {author} {\bibinfo {author} {\bibfnamefont {E.~S.}\ \bibnamefont
  {Battistelli}}, \bibinfo {author} {\bibfnamefont {M.}~\bibnamefont
  {De~Petris}}, \bibinfo {author} {\bibfnamefont {L.}~\bibnamefont {Lamagna}},
  \bibinfo {author} {\bibfnamefont {F.}~\bibnamefont {Melchiorri}}, \bibinfo
  {author} {\bibfnamefont {E.}~\bibnamefont {Palladino}}, \bibinfo {author}
  {\bibfnamefont {G.}~\bibnamefont {Savini}}, \bibinfo {author} {\bibfnamefont
  {A.}~\bibnamefont {Cooray}}, \bibinfo {author} {\bibfnamefont
  {A.}~\bibnamefont {Melchiorri}}, \bibinfo {author} {\bibfnamefont
  {Y.}~\bibnamefont {Rephaeli}}, \ and\ \bibinfo {author} {\bibfnamefont
  {M.}~\bibnamefont {Shimon}},\ }\href {\doibase 10.1086/345589} {\bibfield
  {journal} {\bibinfo  {journal} {The Astrophysical Journal}\ }\textbf
  {\bibinfo {volume} {580}},\ \bibinfo {pages} {L101–L104} (\bibinfo {year}
  {2002})}\BibitemShut {NoStop}%
\bibitem [{\citenamefont {{Bahcall}}\ and\ \citenamefont
  {{Wolf}}(1968)}]{Bahcall1968}%
  \BibitemOpen
  \bibfield  {author} {\bibinfo {author} {\bibfnamefont {J.~N.}\ \bibnamefont
  {{Bahcall}}}\ and\ \bibinfo {author} {\bibfnamefont {R.~A.}\ \bibnamefont
  {{Wolf}}},\ }\href {\doibase 10.1086/149589} {\bibfield  {journal} {\bibinfo
  {journal} {apj}\ }\textbf {\bibinfo {volume} {152}},\ \bibinfo {pages} {701}
  (\bibinfo {year} {1968})}\BibitemShut {NoStop}%
\bibitem [{\citenamefont {Rasmussen}(2003)}]{GP}%
  \BibitemOpen
  \bibfield  {author} {\bibinfo {author} {\bibfnamefont {C.~E.}\ \bibnamefont
  {Rasmussen}},\ }in\ \href@noop {} {\emph {\bibinfo {booktitle} {Summer school
  on machine learning}}}\ (\bibinfo  {publisher} {Springer},\ \bibinfo {year}
  {2003})\ pp.\ \bibinfo {pages} {63--71}\BibitemShut {NoStop}%
\bibitem [{\citenamefont {{Bengaly}}\ \emph {et~al.}(2020)\citenamefont
  {{Bengaly}}, \citenamefont {{Gonzalez}},\ and\ \citenamefont
  {{Alcaniz}}}]{Bengaly20}%
  \BibitemOpen
  \bibfield  {author} {\bibinfo {author} {\bibfnamefont {C.~A.~P.}\
  \bibnamefont {{Bengaly}}}, \bibinfo {author} {\bibfnamefont {J.~E.}\
  \bibnamefont {{Gonzalez}}}, \ and\ \bibinfo {author} {\bibfnamefont {J.~S.}\
  \bibnamefont {{Alcaniz}}},\ }\href {\doibase 10.1140/epjc/s10052-020-08522-6}
  {\bibfield  {journal} {\bibinfo  {journal} {European Physical Journal C}\
  }\textbf {\bibinfo {volume} {80}},\ \bibinfo {eid} {936} (\bibinfo {year}
  {2020})},\ \Eprint {http://arxiv.org/abs/2007.13789} {arXiv:2007.13789
  [astro-ph.CO]} \BibitemShut {NoStop}%
\bibitem [{\citenamefont {{Hurier}}\ \emph {et~al.}(2014)\citenamefont
  {{Hurier}}, \citenamefont {{Aghanim}}, \citenamefont {{Douspis}},\ and\
  \citenamefont {{Pointecouteau}}}]{Hurier14}%
  \BibitemOpen
  \bibfield  {author} {\bibinfo {author} {\bibfnamefont {G.}~\bibnamefont
  {{Hurier}}}, \bibinfo {author} {\bibfnamefont {N.}~\bibnamefont {{Aghanim}}},
  \bibinfo {author} {\bibfnamefont {M.}~\bibnamefont {{Douspis}}}, \ and\
  \bibinfo {author} {\bibfnamefont {E.}~\bibnamefont {{Pointecouteau}}},\
  }\href {\doibase 10.1051/0004-6361/201322632} {\bibfield  {journal} {\bibinfo
   {journal} {aap}\ }\textbf {\bibinfo {volume} {561}},\ \bibinfo {eid} {A143}
  (\bibinfo {year} {2014})},\ \Eprint {http://arxiv.org/abs/1311.4694}
  {arXiv:1311.4694 [astro-ph.CO]} \BibitemShut {NoStop}%
\bibitem [{\citenamefont {Ade}\ \emph {et~al.}(2014)\citenamefont {Ade} \emph
  {et~al.}}]{Planck14}%
  \BibitemOpen
  \bibfield  {author} {\bibinfo {author} {\bibfnamefont {P.~A.~R.}\
  \bibnamefont {Ade}} \emph {et~al.} (\bibinfo {collaboration} {Planck}),\
  }\href {\doibase 10.1051/0004-6361/201321523} {\bibfield  {journal} {\bibinfo
   {journal} {Astron. Astrophys.}\ }\textbf {\bibinfo {volume} {571}},\
  \bibinfo {pages} {A29} (\bibinfo {year} {2014})},\ \Eprint
  {http://arxiv.org/abs/1303.5089} {arXiv:1303.5089 [astro-ph.CO]} \BibitemShut
  {NoStop}%
\bibitem [{\citenamefont {{Reichardt}}(2013)}]{Reichardt13}%
  \BibitemOpen
  \bibfield  {author} {\bibinfo {author} {\bibfnamefont {C.~L.~e.}\
  \bibnamefont {{Reichardt}}},\ }\href {\doibase 10.1088/0004-637X/763/2/127}
  {\bibfield  {journal} {\bibinfo  {journal} {apj}\ }\textbf {\bibinfo {volume}
  {763}},\ \bibinfo {eid} {127} (\bibinfo {year} {2013})},\ \Eprint
  {http://arxiv.org/abs/1203.5775} {arXiv:1203.5775 [astro-ph.CO]} \BibitemShut
  {NoStop}%
\bibitem [{\citenamefont {Seikel}\ \emph {et~al.}(2012)\citenamefont {Seikel},
  \citenamefont {Clarkson},\ and\ \citenamefont {Smith}}]{Seikel_2012}%
  \BibitemOpen
  \bibfield  {author} {\bibinfo {author} {\bibfnamefont {M.}~\bibnamefont
  {Seikel}}, \bibinfo {author} {\bibfnamefont {C.}~\bibnamefont {Clarkson}}, \
  and\ \bibinfo {author} {\bibfnamefont {M.}~\bibnamefont {Smith}},\ }\href
  {\doibase 10.1088/1475-7516/2012/06/036} {\bibfield  {journal} {\bibinfo
  {journal} {Journal of Cosmology and Astroparticle Physics}\ }\textbf
  {\bibinfo {volume} {2012}},\ \bibinfo {pages} {036–036} (\bibinfo {year}
  {2012})}\BibitemShut {NoStop}%
\bibitem [{\citenamefont {Shafieloo}\ \emph {et~al.}(2012)\citenamefont
  {Shafieloo}, \citenamefont {Kim},\ and\ \citenamefont
  {Linder}}]{Shafieloo12}%
  \BibitemOpen
  \bibfield  {author} {\bibinfo {author} {\bibfnamefont {A.}~\bibnamefont
  {Shafieloo}}, \bibinfo {author} {\bibfnamefont {A.~G.}\ \bibnamefont {Kim}},
  \ and\ \bibinfo {author} {\bibfnamefont {E.~V.}\ \bibnamefont {Linder}},\
  }\href {\doibase 10.1103/PhysRevD.85.123530} {\bibfield  {journal} {\bibinfo
  {journal} {Phys. Rev. D}\ }\textbf {\bibinfo {volume} {85}},\ \bibinfo
  {pages} {123530} (\bibinfo {year} {2012})}\BibitemShut {NoStop}%
\bibitem [{\citenamefont {Seikel}\ and\ \citenamefont
  {Clarkson}(2013)}]{seikel13}%
  \BibitemOpen
  \bibfield  {author} {\bibinfo {author} {\bibfnamefont {M.}~\bibnamefont
  {Seikel}}\ and\ \bibinfo {author} {\bibfnamefont {C.}~\bibnamefont
  {Clarkson}},\ }\href {https://arxiv.org/abs/1311.6678} {\enquote {\bibinfo
  {title} {Optimising gaussian processes for reconstructing dark energy
  dynamics from supernovae},}\ } (\bibinfo {year} {2013}),\ \Eprint
  {http://arxiv.org/abs/1311.6678} {arXiv:1311.6678 [astro-ph.CO]} \BibitemShut
  {NoStop}%
\bibitem [{\citenamefont {Zhang}\ and\ \citenamefont {Xia}(2016)}]{Zhang16}%
  \BibitemOpen
  \bibfield  {author} {\bibinfo {author} {\bibfnamefont {M.-J.}\ \bibnamefont
  {Zhang}}\ and\ \bibinfo {author} {\bibfnamefont {J.-Q.}\ \bibnamefont
  {Xia}},\ }\href {\doibase 10.1088/1475-7516/2016/12/005} {\bibfield
  {journal} {\bibinfo  {journal} {Journal of Cosmology and Astroparticle
  Physics}\ }\textbf {\bibinfo {volume} {2016}},\ \bibinfo {pages} {005–005}
  (\bibinfo {year} {2016})}\BibitemShut {NoStop}%
\bibitem [{\citenamefont {Busti}\ \emph {et~al.}(2014)\citenamefont {Busti},
  \citenamefont {Clarkson},\ and\ \citenamefont {Seikel}}]{Busti14}%
  \BibitemOpen
  \bibfield  {author} {\bibinfo {author} {\bibfnamefont {V.~C.}\ \bibnamefont
  {Busti}}, \bibinfo {author} {\bibfnamefont {C.}~\bibnamefont {Clarkson}}, \
  and\ \bibinfo {author} {\bibfnamefont {M.}~\bibnamefont {Seikel}},\ }\href
  {\doibase 10.1093/mnrasl/slu035} {\bibfield  {journal} {\bibinfo  {journal}
  {Monthly Notices of the Royal Astronomical Society: Letters}\ }\textbf
  {\bibinfo {volume} {441}},\ \bibinfo {pages} {L11–L15} (\bibinfo {year}
  {2014})}\BibitemShut {NoStop}%
\bibitem [{\citenamefont {Sahni}\ \emph {et~al.}(2014)\citenamefont {Sahni},
  \citenamefont {Shafieloo},\ and\ \citenamefont {Starobinsky}}]{Sahni14}%
  \BibitemOpen
  \bibfield  {author} {\bibinfo {author} {\bibfnamefont {V.}~\bibnamefont
  {Sahni}}, \bibinfo {author} {\bibfnamefont {A.}~\bibnamefont {Shafieloo}}, \
  and\ \bibinfo {author} {\bibfnamefont {A.~A.}\ \bibnamefont {Starobinsky}},\
  }\href {\doibase 10.1088/2041-8205/793/2/l40} {\bibfield  {journal} {\bibinfo
   {journal} {The Astrophysical Journal}\ }\textbf {\bibinfo {volume} {793}},\
  \bibinfo {pages} {L40} (\bibinfo {year} {2014})}\BibitemShut {NoStop}%
\bibitem [{\citenamefont {Belgacem}\ \emph {et~al.}(2020)\citenamefont
  {Belgacem}, \citenamefont {Foffa}, \citenamefont {Maggiore},\ and\
  \citenamefont {Yang}}]{Belgacem20}%
  \BibitemOpen
  \bibfield  {author} {\bibinfo {author} {\bibfnamefont {E.}~\bibnamefont
  {Belgacem}}, \bibinfo {author} {\bibfnamefont {S.}~\bibnamefont {Foffa}},
  \bibinfo {author} {\bibfnamefont {M.}~\bibnamefont {Maggiore}}, \ and\
  \bibinfo {author} {\bibfnamefont {T.}~\bibnamefont {Yang}},\ }\href {\doibase
  10.1103/physrevd.101.063505} {\bibfield  {journal} {\bibinfo  {journal}
  {Physical Review D}\ }\textbf {\bibinfo {volume} {101}} (\bibinfo {year}
  {2020}),\ 10.1103/physrevd.101.063505}\BibitemShut {NoStop}%
\bibitem [{\citenamefont {Pinho}\ \emph {et~al.}(2018)\citenamefont {Pinho},
  \citenamefont {Casas},\ and\ \citenamefont {Amendola}}]{Pinho18}%
  \BibitemOpen
  \bibfield  {author} {\bibinfo {author} {\bibfnamefont {A.~M.}\ \bibnamefont
  {Pinho}}, \bibinfo {author} {\bibfnamefont {S.}~\bibnamefont {Casas}}, \ and\
  \bibinfo {author} {\bibfnamefont {L.}~\bibnamefont {Amendola}},\ }\href
  {\doibase 10.1088/1475-7516/2018/11/027} {\bibfield  {journal} {\bibinfo
  {journal} {Journal of Cosmology and Astroparticle Physics}\ }\textbf
  {\bibinfo {volume} {2018}},\ \bibinfo {pages} {027–027} (\bibinfo {year}
  {2018})}\BibitemShut {NoStop}%
\bibitem [{\citenamefont {Cai}\ \emph {et~al.}(2017)\citenamefont {Cai},
  \citenamefont {Tamanini},\ and\ \citenamefont {Yang}}]{Cai17}%
  \BibitemOpen
  \bibfield  {author} {\bibinfo {author} {\bibfnamefont {R.-G.}\ \bibnamefont
  {Cai}}, \bibinfo {author} {\bibfnamefont {N.}~\bibnamefont {Tamanini}}, \
  and\ \bibinfo {author} {\bibfnamefont {T.}~\bibnamefont {Yang}},\ }\href
  {\doibase 10.1088/1475-7516/2017/05/031} {\bibfield  {journal} {\bibinfo
  {journal} {Journal of Cosmology and Astroparticle Physics}\ }\textbf
  {\bibinfo {volume} {2017}},\ \bibinfo {pages} {031–031} (\bibinfo {year}
  {2017})}\BibitemShut {NoStop}%
\bibitem [{\citenamefont {Haridasu}\ \emph {et~al.}(2018)\citenamefont
  {Haridasu}, \citenamefont {Luković}, \citenamefont {Moresco},\ and\
  \citenamefont {Vittorio}}]{Haridasu18}%
  \BibitemOpen
  \bibfield  {author} {\bibinfo {author} {\bibfnamefont {B.~S.}\ \bibnamefont
  {Haridasu}}, \bibinfo {author} {\bibfnamefont {V.~V.}\ \bibnamefont
  {Luković}}, \bibinfo {author} {\bibfnamefont {M.}~\bibnamefont {Moresco}}, \
  and\ \bibinfo {author} {\bibfnamefont {N.}~\bibnamefont {Vittorio}},\ }\href
  {\doibase 10.1088/1475-7516/2018/10/015} {\bibfield  {journal} {\bibinfo
  {journal} {Journal of Cosmology and Astroparticle Physics}\ }\textbf
  {\bibinfo {volume} {2018}},\ \bibinfo {pages} {015–015} (\bibinfo {year}
  {2018})}\BibitemShut {NoStop}%
\bibitem [{\citenamefont {Zhang}\ and\ \citenamefont {Li}(2018)}]{Zhang18}%
  \BibitemOpen
  \bibfield  {author} {\bibinfo {author} {\bibfnamefont {M.-J.}\ \bibnamefont
  {Zhang}}\ and\ \bibinfo {author} {\bibfnamefont {H.}~\bibnamefont {Li}},\
  }\href {\doibase 10.1140/epjc/s10052-018-5953-3} {\bibfield  {journal}
  {\bibinfo  {journal} {The European Physical Journal C}\ }\textbf {\bibinfo
  {volume} {78}} (\bibinfo {year} {2018}),\
  10.1140/epjc/s10052-018-5953-3}\BibitemShut {NoStop}%
\bibitem [{\citenamefont {Wang}\ and\ \citenamefont {Meng}(2017)}]{Wang17}%
  \BibitemOpen
  \bibfield  {author} {\bibinfo {author} {\bibfnamefont {D.}~\bibnamefont
  {Wang}}\ and\ \bibinfo {author} {\bibfnamefont {X.-H.}\ \bibnamefont
  {Meng}},\ }\href {\doibase 10.1103/physrevd.95.023508} {\bibfield  {journal}
  {\bibinfo  {journal} {Physical Review D}\ }\textbf {\bibinfo {volume} {95}}
  (\bibinfo {year} {2017}),\ 10.1103/physrevd.95.023508}\BibitemShut {NoStop}%
\bibitem [{\citenamefont {Sharma}\ \emph {et~al.}(2022)\citenamefont {Sharma},
  \citenamefont {Mukherjee},\ and\ \citenamefont {Jassal}}]{Sharma22}%
  \BibitemOpen
  \bibfield  {author} {\bibinfo {author} {\bibfnamefont {R.}~\bibnamefont
  {Sharma}}, \bibinfo {author} {\bibfnamefont {A.}~\bibnamefont {Mukherjee}}, \
  and\ \bibinfo {author} {\bibfnamefont {H.~K.}\ \bibnamefont {Jassal}},\
  }\href {\doibase 10.1140/epjp/s13360-022-02397-0} {\bibfield  {journal}
  {\bibinfo  {journal} {The European Physical Journal Plus}\ }\textbf {\bibinfo
  {volume} {137}} (\bibinfo {year} {2022}),\
  10.1140/epjp/s13360-022-02397-0}\BibitemShut {NoStop}%
\bibitem [{\citenamefont {Bonilla}\ \emph {et~al.}(2021)\citenamefont
  {Bonilla}, \citenamefont {Kumar},\ and\ \citenamefont {Nunes}}]{Bonilla20}%
  \BibitemOpen
  \bibfield  {author} {\bibinfo {author} {\bibfnamefont {A.}~\bibnamefont
  {Bonilla}}, \bibinfo {author} {\bibfnamefont {S.}~\bibnamefont {Kumar}}, \
  and\ \bibinfo {author} {\bibfnamefont {R.~C.}\ \bibnamefont {Nunes}},\ }\href
  {\doibase 10.1140/epjc/s10052-021-08925-z} {\bibfield  {journal} {\bibinfo
  {journal} {Eur. Phys. J. C}\ }\textbf {\bibinfo {volume} {81}},\ \bibinfo
  {pages} {127} (\bibinfo {year} {2021})},\ \Eprint
  {http://arxiv.org/abs/2011.07140} {arXiv:2011.07140 [astro-ph.CO]}
  \BibitemShut {NoStop}%
\bibitem [{\citenamefont {Sabogal}\ \emph {et~al.}(2024)\citenamefont
  {Sabogal}, \citenamefont {Akarsu}, \citenamefont {Bonilla}, \citenamefont
  {Di~Valentino},\ and\ \citenamefont {Nunes}}]{Saboga24}%
  \BibitemOpen
  \bibfield  {author} {\bibinfo {author} {\bibfnamefont {M.~A.}\ \bibnamefont
  {Sabogal}}, \bibinfo {author} {\bibfnamefont {O.}~\bibnamefont {Akarsu}},
  \bibinfo {author} {\bibfnamefont {A.}~\bibnamefont {Bonilla}}, \bibinfo
  {author} {\bibfnamefont {E.}~\bibnamefont {Di~Valentino}}, \ and\ \bibinfo
  {author} {\bibfnamefont {R.~C.}\ \bibnamefont {Nunes}},\ }\href {\doibase
  10.1140/epjc/s10052-024-13081-1} {\bibfield  {journal} {\bibinfo  {journal}
  {Eur. Phys. J. C}\ }\textbf {\bibinfo {volume} {84}},\ \bibinfo {pages} {703}
  (\bibinfo {year} {2024})},\ \Eprint {http://arxiv.org/abs/2407.04223}
  {arXiv:2407.04223 [astro-ph.CO]} \BibitemShut {NoStop}%
\bibitem [{\citenamefont {Oliveira}\ \emph {et~al.}(2024)\citenamefont
  {Oliveira}, \citenamefont {Avila}, \citenamefont {Bernui}, \citenamefont
  {Bonilla},\ and\ \citenamefont {Nunes}}]{Oliveira23}%
  \BibitemOpen
  \bibfield  {author} {\bibinfo {author} {\bibfnamefont {F.}~\bibnamefont
  {Oliveira}}, \bibinfo {author} {\bibfnamefont {F.}~\bibnamefont {Avila}},
  \bibinfo {author} {\bibfnamefont {A.}~\bibnamefont {Bernui}}, \bibinfo
  {author} {\bibfnamefont {A.}~\bibnamefont {Bonilla}}, \ and\ \bibinfo
  {author} {\bibfnamefont {R.~C.}\ \bibnamefont {Nunes}},\ }\href {\doibase
  10.1140/epjc/s10052-024-12953-w} {\bibfield  {journal} {\bibinfo  {journal}
  {Eur. Phys. J. C}\ }\textbf {\bibinfo {volume} {84}},\ \bibinfo {pages} {636}
  (\bibinfo {year} {2024})},\ \Eprint {http://arxiv.org/abs/2311.14216}
  {arXiv:2311.14216 [astro-ph.CO]} \BibitemShut {NoStop}%
\bibitem [{\citenamefont {Avila}\ \emph {et~al.}(2022)\citenamefont {Avila},
  \citenamefont {Bernui}, \citenamefont {Bonilla},\ and\ \citenamefont
  {Nunes}}]{Avila22}%
  \BibitemOpen
  \bibfield  {author} {\bibinfo {author} {\bibfnamefont {F.}~\bibnamefont
  {Avila}}, \bibinfo {author} {\bibfnamefont {A.}~\bibnamefont {Bernui}},
  \bibinfo {author} {\bibfnamefont {A.}~\bibnamefont {Bonilla}}, \ and\
  \bibinfo {author} {\bibfnamefont {R.~C.}\ \bibnamefont {Nunes}},\ }\href
  {\doibase 10.1140/epjc/s10052-022-10561-0} {\bibfield  {journal} {\bibinfo
  {journal} {Eur. Phys. J. C}\ }\textbf {\bibinfo {volume} {82}},\ \bibinfo
  {pages} {594} (\bibinfo {year} {2022})},\ \Eprint
  {http://arxiv.org/abs/2201.07829} {arXiv:2201.07829 [astro-ph.CO]}
  \BibitemShut {NoStop}%
\bibitem [{\citenamefont {{Avila}}\ \emph {et~al.}(2022)\citenamefont
  {{Avila}}, \citenamefont {{Bernui}}, \citenamefont {{Nunes}}, \citenamefont
  {{de Carvalho}},\ and\ \citenamefont {{Novaes}}}]{Avila22a}%
  \BibitemOpen
  \bibfield  {author} {\bibinfo {author} {\bibfnamefont {F.}~\bibnamefont
  {{Avila}}}, \bibinfo {author} {\bibfnamefont {A.}~\bibnamefont {{Bernui}}},
  \bibinfo {author} {\bibfnamefont {R.~C.}\ \bibnamefont {{Nunes}}}, \bibinfo
  {author} {\bibfnamefont {E.}~\bibnamefont {{de Carvalho}}}, \ and\ \bibinfo
  {author} {\bibfnamefont {C.~P.}\ \bibnamefont {{Novaes}}},\ }\href {\doibase
  10.1093/mnras/stab3122} {\bibfield  {journal} {\bibinfo  {journal} {mnras}\
  }\textbf {\bibinfo {volume} {509}},\ \bibinfo {pages} {2994} (\bibinfo {year}
  {2022})},\ \Eprint {http://arxiv.org/abs/2111.08541} {arXiv:2111.08541
  [astro-ph.CO]} \BibitemShut {NoStop}%
\bibitem [{\citenamefont {Bonilla}\ \emph {et~al.}(2022)\citenamefont
  {Bonilla}, \citenamefont {Kumar}, \citenamefont {Nunes},\ and\ \citenamefont
  {Pan}}]{Bonilla21}%
  \BibitemOpen
  \bibfield  {author} {\bibinfo {author} {\bibfnamefont {A.}~\bibnamefont
  {Bonilla}}, \bibinfo {author} {\bibfnamefont {S.}~\bibnamefont {Kumar}},
  \bibinfo {author} {\bibfnamefont {R.~C.}\ \bibnamefont {Nunes}}, \ and\
  \bibinfo {author} {\bibfnamefont {S.}~\bibnamefont {Pan}},\ }\href {\doibase
  10.1093/mnras/stac687} {\bibfield  {journal} {\bibinfo  {journal} {Mon. Not.
  Roy. Astron. Soc.}\ }\textbf {\bibinfo {volume} {512}},\ \bibinfo {pages}
  {4231} (\bibinfo {year} {2022})},\ \Eprint {http://arxiv.org/abs/2102.06149}
  {arXiv:2102.06149 [astro-ph.CO]} \BibitemShut {NoStop}%
\bibitem [{\citenamefont {Rasanen}\ \emph {et~al.}(2016)\citenamefont
  {Rasanen}, \citenamefont {Valiviita},\ and\ \citenamefont
  {Kosonen}}]{Rasanen:2015kca}%
  \BibitemOpen
  \bibfield  {author} {\bibinfo {author} {\bibfnamefont {S.}~\bibnamefont
  {Rasanen}}, \bibinfo {author} {\bibfnamefont {J.}~\bibnamefont {Valiviita}},
  \ and\ \bibinfo {author} {\bibfnamefont {V.}~\bibnamefont {Kosonen}},\ }\href
  {\doibase 10.1088/1475-7516/2016/04/050} {\bibfield  {journal} {\bibinfo
  {journal} {JCAP}\ }\textbf {\bibinfo {volume} {04}},\ \bibinfo {pages} {050}
  (\bibinfo {year} {2016})},\ \Eprint {http://arxiv.org/abs/1512.05346}
  {arXiv:1512.05346 [astro-ph.CO]} \BibitemShut {NoStop}%
\bibitem [{\citenamefont {Kumar}\ \emph {et~al.}(2018)\citenamefont {Kumar},
  \citenamefont {Nunes},\ and\ \citenamefont {Yadav}}]{Kumar:2018yhh}%
  \BibitemOpen
  \bibfield  {author} {\bibinfo {author} {\bibfnamefont {S.}~\bibnamefont
  {Kumar}}, \bibinfo {author} {\bibfnamefont {R.~C.}\ \bibnamefont {Nunes}}, \
  and\ \bibinfo {author} {\bibfnamefont {S.~K.}\ \bibnamefont {Yadav}},\ }\href
  {\doibase 10.1103/PhysRevD.98.043521} {\bibfield  {journal} {\bibinfo
  {journal} {Phys. Rev. D}\ }\textbf {\bibinfo {volume} {98}},\ \bibinfo
  {pages} {043521} (\bibinfo {year} {2018})},\ \Eprint
  {http://arxiv.org/abs/1803.10229} {arXiv:1803.10229 [astro-ph.CO]}
  \BibitemShut {NoStop}%
\bibitem [{\citenamefont {{D'Agostino}}\ and\ \citenamefont
  {{Nunes}}(2023)}]{Rocco23}%
  \BibitemOpen
  \bibfield  {author} {\bibinfo {author} {\bibfnamefont {R.}~\bibnamefont
  {{D'Agostino}}}\ and\ \bibinfo {author} {\bibfnamefont {R.~C.}\ \bibnamefont
  {{Nunes}}},\ }\href {\doibase 10.1103/PhysRevD.108.023523} {\bibfield
  {journal} {\bibinfo  {journal} {prd}\ }\textbf {\bibinfo {volume} {108}},\
  \bibinfo {eid} {023523} (\bibinfo {year} {2023})},\ \Eprint
  {http://arxiv.org/abs/2307.13464} {arXiv:2307.13464 [astro-ph.CO]}
  \BibitemShut {NoStop}%
\end{thebibliography}%

\end{document}